\documentclass[5p]{elsarticle}

\usepackage{amsmath}
\usepackage{amssymb}
\usepackage{color}
\usepackage{algorithm}
\usepackage{algpseudocode}
\usepackage{natbib}
\usepackage{subfig}
\usepackage{hyperref}
\usepackage[all]{hypcap}

\hypersetup{pdfborder={0 0 0},
	colorlinks=true,
	linkcolor=black,
	citecolor=royalblue,
	urlcolor=blue}

\journal{Astronomy and Computing}
\biboptions{authoryear}

\begin{document}

\begin{frontmatter}

\title{High Performance Computing for gravitational lens modeling:\\ single vs double precision on GPUs and CPUs}

\author[LastroAddress]{Markus Rexroth}
\author[LastroAddress]{Christoph Sch\"{a}fer\corref{maincorrespondingauthor}}

\cortext[maincorrespondingauthor]{Corresponding author}
\ead{christophernstrerne.schaefer@epfl.ch}

\author[ScitasAddress]{Gilles Fourestey}
\author[LastroAddress,LamAddress]{Jean-Paul Kneib}

\address[LastroAddress]{Institute of Physics, Laboratory of Astrophysics, Ecole Polytechnique F\'{e}d\'{e}rale de Lausanne (EPFL), Observatoire de Sauverny,\\ 1290 Versoix, Switzerland}
\address[ScitasAddress]{SCITAS, Ecole Polytechnique F\'{e}d\'{e}rale de Lausanne (EPFL), 1015 Lausanne, Switzerland}
\address[LamAddress]{Aix Marseille Universit\'{e}, CNRS, LAM (Laboratoire d'Astrophysique de Marseille) UMR 7326, 13388, Marseille, France}

\begin{abstract}
Strong gravitational lensing is a powerful probe of cosmology and the dark matter distribution. Efficient lensing software is already a necessity to fully use its potential and the performance demands will only increase with the upcoming generation of telescopes. In this paper, we study the possible impact of High Performance Computing techniques on a performance-critical part of the widely used lens modeling software \texttt{LENSTOOL}. We implement the algorithm once as a highly optimized CPU version and once with graphics card acceleration for a simple parametric lens model. In addition, we study the impact of finite machine precision on the lensing algorithm. While double precision is the default choice for scientific applications, we find that single precision can be sufficiently accurate for our purposes and lead to a big speedup. Therefore we develop and present a mixed precision algorithm which only uses double precision when necessary. We measure the performance of the different implementations and find that the use of High Performance Computing Techniques dramatically improves the code performance both on CPUs and GPUs. Compared to the current \texttt{LENSTOOL} implementation on 12 CPU cores, we obtain speedup factors of up to 170. We achieve this optimal performance by using our mixed precision algorithm on a high-end GPU which is common in modern supercomputers. We also show that these techniques reduce the energy consumption by up to 98\%. Furthermore, we demonstrate that a highly competitive speedup can be reached with consumer GPUs. While they are an order of magnitude cheaper than the high-end graphics cards, they are rarely used for scientific computations due to their low double precision performance. However, our mixed precision algorithm unlocks their full potential. Consequently, the consumer GPU delivers a speedup which is only a factor of four lower than the best speedup achieved by a high-end GPU.
\end{abstract}


\end{frontmatter}

\section{Introduction} 
The $\Lambda$CDM cosmology standard model describes our universe with great precision, but it also introduces two unknown quantities, Dark Energy and Dark Matter. They dominate the energy density of the universe \citep[e.g.,][]{PlanckCollaboration2016}, but their physical nature has so far remained elusive. Consequently their study is one of the prime targets of cosmological research. \\
\\
Strong gravitational lensing is a unique tool for cosmology, as it is sensitive to the total (baryonic and dark) matter density and thus it probes Dark Matter directly \citep[see e.g.][for reviews]{Kneib2011,SaasFee2006}. Its application has led to constraints on cosmological parameters \citep[e.g.,][]{Jullo2010,Bonvin2017} and the Dark Matter self-interaction cross-section \citep[e.g.,][]{Randall2008,Bradac2008}. In addition, the magnification effect of a strong gravitational lens can be used to study the early universe and to constrain its reionization \citep[e.g.,][]{Atek2015a,Ishigaki2015}. Strong lensing requires deep, high-resolution data and indeed the field has prospered thanks to programs like the Cluster Lensing And Supernova survey with \textit{Hubble} \citep[CLASH,][]{Postman2012} and the \textit{Hubble} Frontier Fields \citep[HFF,][]{Lotz2017}.\\
\\
Future missions like Euclid, the \textit{James Webb Space Telescope (JWST)}, the Large Synoptic Survey Telescope (LSST), and the \textit{Wide Field Infrared Survey Telescope (WFIRST)} will provide a large amount of excellent data sets for lensing. These will enable the lensing community to further push the boundaries of cosmological knowledge. This, however, will only be feasible if we are able to efficiently harvest the wealth of information available in the data. This will be a challenge, e.g. due to the amount of data available or the high quality of the data, which permits the creation of lens models with a high level of detail and precision, but also requires more computing time. Gravitational lensing software and pipelines will have to be ready to process these data sets in a reasonable amount of time.\\
\\
Therefore we are currently redesigning the strong lensing software \texttt{LENSTOOL}\footnote{Open source software publicly available at \url{https://projets.lam.fr/projects/lenstool/wiki}} \citep{Jullo2007,Kneib1996}. \texttt{LENSTOOL} has been successfully used to model many strong lensing galaxy clusters with high precision \citep[see e.g.][for recent lens models]{Jauzac2014,Jauzac2015b,Limousin2016} and has been serving the lensing community for more than two decades. In a recent comparison of strong lensing modeling software it has performed very well \citep{Meneghetti2016}. However, the HFF data sets provided the greatest number of lensing constraints so far and this posed a computing challenge for \texttt{LENSTOOL}. It took several weeks to compute a single HFF lens model and several different lens models from different priors are required to find the best fitting model. \\
\\
The new version is designed to meet this computation challenge by using High Performance Computing (HPC) methods. The \texttt{LENSTOOL} algorithms are very well suited for massive parallelism and we employ this technique to accelerate the computations. While we focus on lensing by galaxy clusters, a recent publication by \citet{Tessore2016} has shown that massive parallelism holds also great promise for the modeling of galaxy lenses. In this paper, we discuss the central lensing algorithm of \texttt{LENSTOOL} and in particular the performance-critical computation of deflection potential gradients. We have implemented the gradient computation algorithm using two different hardware types in order to be able to compare performance. The first version is a highly optimized and parallelized CPU code and the second version uses Graphics Processing Unit (GPU) acceleration. \\
\\
During the development phase, we have asked ourselves the question: Can we do even better by using single precision instead of the commonly used double precision? The computing power of both CPUs and GPUs is higher for single precision \citep[see e.g.][]{Eijkhout2016,Besl2013}, so we can expect a significant performance improvement. The downside is that this might lead to an error in our results if single precision is not precise enough for our computations. Therefore we use error propagation to compute the impact of single precision on the results of the central lensing algorithm. In addition, we measure and compare the single and double precision performance of both CPU and GPU implementations. \\
\\
The paper is organized as follows: Section~\ref{section_strong_lensing} gives a concise introduction to strong gravitational lensing and the \texttt{LENSTOOL} algorithms. It also presents the CPU and GPU implementations. Section~\ref{section_finite_precision_errors} introduces the single and double precision floating-point representations and investigates if single precision is precise enough for our computations. We present and compare the performance measurements of the single and double precision CPU and GPU implementations in section~\ref{section_benchmarks}. We discuss our results in section~\ref{section_discussion} and conclude in section~\ref{section_conclusion}.

\section{Accelerating lensing with massive parallelism}\label{section_strong_lensing}
\subsection{Strong gravitational lensing}\label{subsection_strong_gravitational_lensing}
Galaxies and galaxy clusters are so dense that they locally deform space-time. As a result, they can act as a lens for background objects, which are magnified and distorted or even multiply imaged. Lensing also changes the locations at which we observe the lensed images on the sky so that they are typically not coincident with the locations at which we would observe the background sources in the absence of lensing. In practice, we can only observe the lensed images of a background source, but not the background source itself. However, the position of the background source on the sky can be calculated with the lens equation \citep[see e.g. the reviews][for a derivation]{Kneib2011,Bartelmann2001},
\begin{equation}
	\boldsymbol{\beta} = \boldsymbol{\theta} - \boldsymbol{\alpha}(\boldsymbol{\theta}),\label{equation:lens_equation}
\end{equation}
where the two dimensional vectors $\boldsymbol{\beta}$, $\boldsymbol{\theta}$, and $\boldsymbol{\alpha}$ describe respectively the location of the source in the source plane, the location of the lensed image in the image plane, and the scaled deflection angle. Note that these quantities are angles. In the case of multiple images, the lens equation has more than one solution $\boldsymbol{\theta}$ for a fixed value of $\boldsymbol{\beta}$ \citep[e.g.,][]{Bartelmann2001}. The lens equation is derived under the assumption that we have only one lens, that the gravitational field is weak enough so that the field equations of General Relativity can be linearized, that we can use the Born approximation, and that the physical extent of the lens is small compared to the angular diameter distances between observer and lens, $D_{\text{OL}}$, and lens and source, $D_{\text{LS}}$. \\
\\
The background objects are typically extended sources like galaxies. The shape of the lensed images will differ from the shape of the source, since the light coming from the object at coordinate $\boldsymbol{\beta}^{\prime}$ will be lensed slightly differently than the light coming from the object at coordinate $\boldsymbol{\beta}^{\prime\prime}$ \citep[e.g.,][]{Bartelmann2001}. Therefore we can use the lens equation to compute $\boldsymbol{\theta}$ for each coordinate $\boldsymbol{\beta}$ of the object and thus the shape of the lensed image due to distortion and magnification.\\
\\
The scaled deflection angle $\boldsymbol{\alpha}$ is the gradient of the deflection potential $\psi$,
\begin{align}
	&\boldsymbol{\alpha} = \nabla \psi, \label{equation:alpha_gradient_psi}\\
	&\psi(\boldsymbol{\theta}) = \frac{1}{\pi} \int_{\mathbb{R}^2}~\text{d}^2\theta^{\prime}~\kappa(\boldsymbol{\theta^{\prime}}) \ln|\boldsymbol{\theta} - \boldsymbol{\theta^{\prime}}|, \label{equation:psi_integral_kappa}
\end{align}
and $\psi$ depends on the dimensionless projected surface mass density $\kappa$,
\begin{align}
	&\kappa(\boldsymbol{\theta}) = \frac{\Sigma(\boldsymbol{\theta})}{\Sigma_{\text{crit}}},\\
	&\Sigma_{\text{crit}} = \frac{c^2}{4\pi G} \frac{D_{\text{OS}}}{D_{\text{OL}} D_{\text{LS}}}, 
\end{align} 
where $\Sigma(\boldsymbol{\theta})$ is the projected surface mass density,
\begin{equation}
	\Sigma(\boldsymbol{\theta}) = \int~\text{d}z~\rho(\boldsymbol{\theta},z),
\end{equation} 
and we defined the critical projected surface mass density $\Sigma_{\text{crit}}$. Here $\rho$ is the mass density, $c$ is the speed of light, $G$ is the gravitational constant, and $D_{\text{OS}}$ is the angular diameter distance between observer and source. We can see from these equations that $\boldsymbol{\alpha}$ and thus the strength of the lensing effect depend on the projected surface mass density. Therefore lensing probes the total surface mass density of the lens, including baryonic and Dark Matter components.\\
\\
The value of $\kappa$ is a good indicator to distinguish the so-called ``weak'' and ``strong'' lensing regimes. In the case of weak lensing, the lensed image appears slightly magnified and distorted, and in the case of strong lensing, the image is strongly magnified and distorted and multiple images appear. A mass distribution which has $\kappa \geq 1$ somewhere produces multiple images for some source positions $\boldsymbol{\beta}$ \citep[e.g.,][]{Bartelmann2001}. In the case of cluster lensing, the strong lensing area and thus the multiple images are typically located in the central regions of the cluster, where the projected surface mass density is large enough \citep[e.g.,][]{Kneib2011}. 
\\
\\
We will illustrate gravitational lensing with an example. We will look at a simple lens model, the Singular Isothermal Sphere (SIS), which we will use for the remainder of this paper as it has a relatively simple mathematical expression and is thus very instructive. The projected surface mass density is
\begin{equation}
	\Sigma(\boldsymbol{\theta}) = \frac{\sigma_{v}^2}{2 G D_{\text{OL}} |\boldsymbol{\theta}|},
\end{equation}
where $\sigma_{v}$ is the line-of-sight velocity dispersion of the ``particles'' (e.g. galaxies in a galaxy cluster), which are assumed to be in virial equilibrium \citep[e.g.,][]{Bartelmann2001}. Thus we have
\begin{align}
	&\kappa(\boldsymbol{\theta}) = \frac{\theta_{E}}{2|\boldsymbol{\theta}|},  \label{equation:kappa_SIS}\\
	&\theta_{\text{E}} = 4\pi \Big(\frac{\sigma_{v}}{c}\Big)^2 \frac{D_{\text{LS}}}{D_{\text{OS}}}, \label{equation:Einstein_angle_SIS}
\end{align}
where we defined the Einstein deflection angle $\theta_{\text{E}}$. Using equations~\ref{equation:alpha_gradient_psi} and \ref{equation:psi_integral_kappa}, we find that the magnitude of the scaled deflection angle is constant, 
\begin{equation}
	|\boldsymbol{\alpha}| = \theta_{\text{E}}. \label{equation:Magnitude_alpha_SIS}
\end{equation}
We see that the lens equation has infinitely many solutions for $\boldsymbol{\beta} = \mathbf{0}$, namely each point on the circle with radius $\theta_{\text{E}}$. Therefore a background source at this location will be strongly lensed into a perfect Einstein ring.

\subsection{Strong lensing algorithm}
\subsubsection{Overview}
\begin{figure}
\begin{center}
\includegraphics[width=0.48\textwidth]{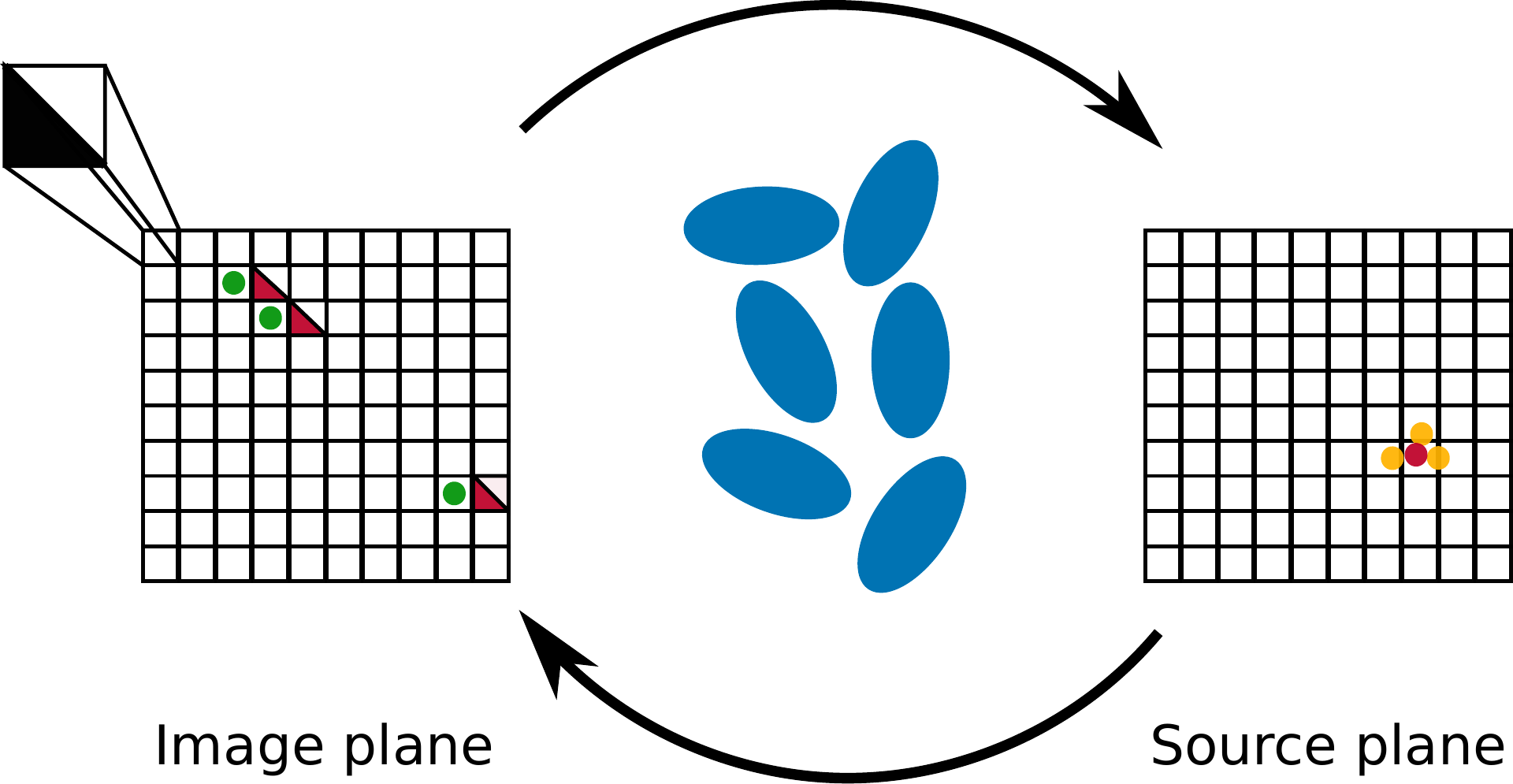}
\end{center}
\caption{Lenstool computes the multiple image positions predicted by a lens model (red triangles, image plane). In the first step, it lenses the observed multiple images (green dots, image plane) onto their respective predicted sources (yellow dots, source plane) and computes their barycenter (red dot, source plane). In the second step, it decomposes the image plane pixels into triangles and lenses each triangle into the source plane. Every time that the source plane triangle includes the barycenter, a predicted multiple image is found. If the lens model is close to the true model, these re-lensed images will be located very close to the observed images. Note that image plane pixels lensed into the source plane will typically be distorted due to the strong lensing effect. We do not show this effect to keep the figure simple. As a result of this distortion, squares are not always mapped onto squares and we thus have to partition the pixels into triangles (top left corner, image plane), which are always mapped onto triangles.  }\label{figure_lenstool_relensing}
\end{figure}

\texttt{LENSTOOL} models strong lensing galaxy clusters by using parametric models of the large-scale cluster halos and the galaxy-scale halos. In a typical merging cluster, we have two large-scale halos and hundreds of galaxy halos. Depending on the chosen parametric model, we have several free parameters such as $x$ and $y$ position, velocity dispersion, etc. for each halo. It is possible to constrain the range of the free parameters or to reduce their number, e.g. by assuming a scaling relation like the Faber-Jackson relation \citep{Faber1976} for galaxy-scale halos \citep{Natarajan1998}. Nevertheless, the best lens model will still be hidden in a massive, high dimensional parameter space. \texttt{LENSTOOL} uses BayeSys3\footnote{Publicly available at \url{http://www.inference.org.uk/bayesys/}}, a Bayesian Markov Chain Monte Carlo (MCMC) software package, to sample this parameter space \citep[see][for a detailed description]{Jullo2007}. For each parameter combination probed by the MCMC, \texttt{LENSTOOL} computes the goodness of fit of the corresponding lens model given the observational data. It does this by modeling the lens with the given set of parameters and, using this model, lensing the observed multiple images into the source plane and subsequently back into the image plane, see figure~\ref{figure_lenstool_relensing}. If the probed lens model is close to the true matter distribution, the re-lensed multiple image positions will be close to the observed multiple image positions and the goodness of fit parameter
\begin{equation}
	\chi^2 = \sum_i \sum_j \frac{(x_{\text{obs}, ij} - x_{ij})^2}{\sigma^2_{ij}}\label{equation:chi2_computation}
\end{equation} 
will be small \citep{Jullo2007}. We denote the observed position of multiple image $j$ of multiple image system $i$ with $x_{\text{obs}, ij}$, the re-lensed position with $x_{ij}$, and the error budget of the position with $\sigma_{ij}$. Since the parameter space probed by the MCMC is massive, it typically takes several weeks of computation time to find the best model for lenses with HFF-like data.\\

There are two ways to speed up the computation. The first is to speed up the MCMC, e.g. by parallelizing it. The second way is to speed up the $\chi^2$ computation. In this paper, we will focus on accelerating a crucial part of it, the gradient computation. Since we will have to take a very precise look at the algorithm when we compute the impact of single and double precision on its result, we will now present a detailed description. 

\subsubsection{Gradient computation in the $\chi^2$ algorithm}
Before we present the $\chi^2$ algorithm, we reformulate the lens equation~\ref{equation:lens_equation} by introducing
\begin{equation}
	\Psi = \frac{D_{\text{OS}}}{D_{\text{LS}}} \psi \label{equation:Modified_deflection_potential}
\end{equation}
and making the gradient dependence explicit:
\begin{align}
	&\beta_{1} = \theta_{1} - \frac{D_{\text{LS}}}{D_{\text{OS}}} (\nabla \Psi(\boldsymbol{\theta}))_{1}, \nonumber \\
	&\beta_{2} = \theta_{2} - \frac{D_{\text{LS}}}{D_{\text{OS}}} (\nabla \Psi(\boldsymbol{\theta}))_{2}.\label{equation:lens_equation_reformulated}
\end{align}
As a result, we only have to compute the constant $D_{\text{LS}}/D_{\text{OS}}$ once instead of for every image pixel. Note that the deflection potential at position $\boldsymbol{\theta}$ is a superposition of all cluster-scale and galaxy-scale deflection potentials $\psi_{\text{cluster}}$ and $\psi_{\text{galaxy}}$,
\begin{equation}
	\psi(\boldsymbol{\theta}) = \sum \psi_{\text{cluster}}(\boldsymbol{\theta}) + \sum \psi_{\text{galaxy}}(\boldsymbol{\theta}),
\end{equation}
\citep[see e.g.][]{Jullo2007} and as a result we have
\begin{align}
	\nabla \Psi(\boldsymbol{\theta}) = \sum \nabla \Psi_{\text{cluster}}(\boldsymbol{\theta}) + \sum \nabla \Psi_{\text{galaxy}}(\boldsymbol{\theta}).
\end{align}
We see that the lens equation is computationally cheap to evaluate once the total gradient $\nabla \Psi$ is known. The computation of $\nabla \Psi$, however, involves potentially complicated gradient calculations for hundreds of potentials and as we will see in the next paragraph, it has to be computed for every pixel in our image. The \textit{Hubble Space Telescope Advanced Camera for Surveys} (HST ACS) produces images with $4096 \times 4096$ pixels at a pixel scale of $\approx0.05~\text{arcsec/pixel}$ \citep{Avila2017}, which we can typically upsample to 0.03~arcsec/pixel \citep{Lotz2017}, so that HFF images have a total of $\approx 6730 \times 6730~\text{pixels} \approx 45~\text{million~pixels}$. This shows that the computation of $\nabla \Psi$ is computationally expensive and an excellent target for speedup with HPC parallelism methods.  \\
\\
The $\chi^2$ computation is now performed as follows. We compute $\nabla \Psi$ for each pixel of the image plane. Then we loop over each multiple image $j$ in each multiple image system $i$. For each multiple image, we use equation~\ref{equation:lens_equation_reformulated} to compute the source coordinates, $\beta_{ij,1}$ and $\beta_{ij,2}$.  Subsequently, we determine the barycenter of the sources of a given multiple image system $i$. If we are close to the true lens model, all multiple images will be mapped onto approximately the same source location, but in general the locations of the predicted sources can differ substantially, which makes it necessary to use the barycenter. In the next step, we re-lens the barycenter back into the image plane to obtain the locations of the multiple images predicted by the lens model. However, the lens equation cannot easily be inverted, so we have to find the locations in a different way. First, we divide each pixel in the image plane into two triangles, see figure~\ref{figure_lenstool_relensing}. We do this because lensing always maps triangles onto triangles, but not squares onto squares. Second, we lens each triangle into the source plane by using equation~\ref{equation:lens_equation_reformulated} and we check if the barycenter is inside this triangle in the source plane. If it is, a predicted multiple image location in the image plane is found. Once we have found the locations of all predicted multiple images for all multiple image systems, we compute the $\chi^2$ according to equation~\ref{equation:chi2_computation}.\\
\\
The gradient calculations will naturally differ for different chosen parametric models. As an example, we present the gradient computation for a generalized form of the SIS, the pseudo-elliptical SIS (henceforth called SIE), in algorithm~\ref{algorithm:gradient_computation}. It is necessary to generalize the parametric model, as we want to use this algorithm to model any SIS lens configuration by simply choosing the appropriate number of lenses and parameter values. We expand our treatment of the SIS in subsection~\ref{subsection_strong_gravitational_lensing} by following the procedure in \citet{Golse2002a}. We introduce the pseudo-ellipticity of the deflection potential, $\epsilon$, and the coordinate system
\begin{align}
	&R = \sqrt{\theta_{1,\epsilon}^{2} + \theta_{2,\epsilon}^{2}}, \nonumber \\
	&\phi = \arctan\Big(\frac{\theta_{2,\epsilon}}{\theta_{1,\epsilon}} \Big),
\end{align}
with
\begin{align}
	&\theta_{1,\epsilon} = \sqrt{a_{1,\epsilon}}~\theta_{1}, \nonumber \\
    &\theta_{2,\epsilon} = \sqrt{a_{2,\epsilon}}~\theta_{2},\\
	&a_{1,\epsilon} = 1 - \epsilon, \nonumber \\
	&a_{2,\epsilon} = 1 + \epsilon.
\end{align}
Note that we call $\epsilon$ a pseudo-ellipticity, because the resulting elliptical shapes will only correspond to ellipses with classical ellipticty $\epsilon^{\prime} = 1 - b/a$, where $a$ and $b$ are the semi-major and semi-minor axes of the ellipse, for small values of $\epsilon$ \citep{Golse2002a}. Therefore we assume in the following $\epsilon \ll 1$. The advantage of using a pseudo-elliptical parametric model is that it leads to relatively simple analytic expressions of the derived lensing quantities \citep{Golse2002a}. Now we can simply calculate the values of the pseudo-elliptical deflection potential $\psi_{\epsilon}$ at location $\boldsymbol{\theta}$ by using the relation \citep{Golse2002a}
\begin{equation}
	\psi_{\epsilon}(\boldsymbol{\theta}) = \psi(R,\phi),
\end{equation}
and analogous for $\Psi_{\epsilon}$. The resulting pseudo-elliptical shape is stretched along the $\theta_{1}$-axis, so that we have $\Phi = 0$, where $\Phi$ is the counter-clockwise angle between the semi-major-axis and the $\theta_{1}$-axis. Algorithm~\ref{algorithm:gradient_computation} extends this approach to potentials with $\Phi \ne 0$ by using rotations. We obtain the following equations for the scaled deflection angle \citep{Golse2002a},
\begin{align}
	&\alpha_{1,\epsilon}(\boldsymbol{\theta}) = |\boldsymbol{\alpha}(R,\phi)|~\sqrt{a_{1,\epsilon}}~\cos(\phi), \nonumber \\
	&\alpha_{2,\epsilon}(\boldsymbol{\theta}) = |\boldsymbol{\alpha}(R,\phi)|~\sqrt{a_{2,\epsilon}}~\sin(\phi). \label{equation:alpha_pseudo-elliptical}
\end{align}
We can now combine the equations~\ref{equation:alpha_gradient_psi}, \ref{equation:Einstein_angle_SIS}, \ref{equation:Magnitude_alpha_SIS}, \ref{equation:Modified_deflection_potential}, and \ref{equation:alpha_pseudo-elliptical} to derive the gradient expressions for the SIE,
\begin{align}
	& \big(\nabla \Psi_{\epsilon} \big)_{1} = (1-\epsilon)~b_{0}~\frac{\theta_{1}}{R}, \nonumber \\
	& \big(\nabla \Psi_{\epsilon} \big)_{2} = (1+\epsilon)~b_{0}~\frac{\theta_{2}}{R},
\end{align}
where we introduced the parameter
\begin{equation}
	b_{0} = 4\pi \Big(\frac{\sigma_{v}}{c}\Big)^{2}.
\end{equation}
The presented equations for the SIE always reduce to the previously presented equations for the spherical SIS for $\epsilon = 0$.

\begin{algorithm}
	\renewcommand{\hypcapspace}{30pt} 
	\capstart 
	\renewcommand{\hypcapspace}{0.5\baselineskip} 
	\begin{algorithmic}[1]
		\Statex \textbf{Require:} $\boldsymbol{\theta_{\text{center}}}$, $b_{0}$, $\epsilon$, $\Phi$ $\forall$ SIE lenses, image \textbf{\textsf{I}}
		\Statex \textbf{Output:} $\nabla \Psi_{\epsilon}$ $\forall$ pixels $(\theta_{1},\theta_{2}) \in$ \textbf{\textsf{I}}
		\State \textbf{Procedure} \textsc{gradient}(\textbf{\textsf{I}},$\{\boldsymbol{\theta}_{\text{center},i},b_{0,i},\epsilon_{i},\Phi_{i}\}$)\textbf{:}
		\ForAll{$(\theta_{1},\theta_{2}) \in$ \textbf{\textsf{I}}}
		\ForAll{SIE lenses}
		\State $\Delta \theta_{1,i} \gets \theta_{1} - \theta_{\text{center},i,1}$
		\State $\Delta \theta_{2,i} \gets \theta_{2} - \theta_{\text{center},i,2}$
		\State $\Delta \theta_{1,i}^{\prime} \gets \Delta \theta_{1,i} \cos(\Phi_{i}) + \Delta \theta_{2,i} \sin(\Phi_{i})$
		\State $\Delta \theta_{2,i}^{\prime} \gets \Delta \theta_{2,i} \cos(\Phi_{i}) - \Delta \theta_{1,i} \sin(\Phi_{i})$
		\State $R_{i} \gets \textbf{sqrt}\big((\Delta \theta_{1,i}^{\prime})^{2} (1-\epsilon_{i}) + (\Delta \theta_{2,i}^{\prime})^{2} (1+ \epsilon_{i}) \big)$
		\State $(\nabla \Psi_{\epsilon})_{1,i} \gets (1- \epsilon_{i})~b_{0,i}~ \Delta \theta_{1,i}^{\prime}/R_{i}$
		\State $(\nabla \Psi_{\epsilon})_{2,i} \gets (1+ \epsilon_{i})~b_{0,i}~ \Delta \theta_{2,i}^{\prime}/R_{i}$
		\State $(\nabla \Psi_{\epsilon})_{1,i}^{\prime} \gets (\nabla \Psi_{\epsilon})_{1,i} \cos(-\Phi_{i})$
		\Statex $\hphantom{alignmentAlignm} + (\nabla \Psi_{\epsilon})_{2,i} \sin(-\Phi_{i})$
		\State $(\nabla \Psi_{\epsilon})_{2,i}^{\prime} \gets (\nabla \Psi_{\epsilon})_{2,i} \cos(-\Phi_{i})$
		\Statex $\hphantom{alignmentAlignm} - (\nabla \Psi_{\epsilon})_{1,i} \sin(-\Phi_{i})$
		\EndFor
		\State $(\nabla \Psi_{\epsilon})_{1} \gets \sum_i~(\nabla \Psi_{\epsilon})_{1,i}^{\prime}$
		\State $(\nabla \Psi_{\epsilon})_{2} \gets \sum_i~(\nabla \Psi_{\epsilon})_{2,i}^{\prime}$
		\EndFor
		\State \Return $\{\nabla \Psi_{\epsilon}\}$
	\end{algorithmic}
	\caption{Compute $\nabla \Psi_{\epsilon}$ in each image pixel for a SIE}\label{algorithm:gradient_computation}
\end{algorithm}

\subsection{CPU and GPU implementations}
\begin{figure}
	\begin{center}
		\includegraphics[width=0.48\textwidth]{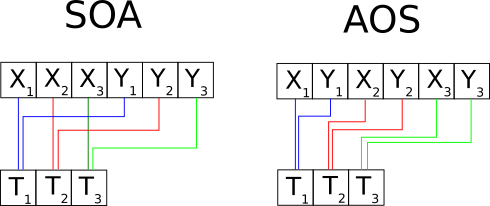}
	\end{center}
	\caption{Illustration of Structures of Arrays (SoA) and Arrays of Structures (AoS) formats. The three data sets T consist each of parameters X and Y. In the SoA format, the data is stored in arrays of the parameters X and Y, so the X parameter of T$_{1}$ is directly followed by the X parameter of T$_{2}$. In the AoS format, the data is stored in data structures, so the X parameter of T$_{1}$ is directly followed by the Y parameter of T$_{1}$. In the SoA format, the same parameters of the different data sets are thus stored in contiguous parts of the memory, which is usually a benefit for vectorized computations. }\label{figure_SOA}
\end{figure}

\begin{figure*}
	\begin{center}
		\includegraphics[width=0.8\textwidth]{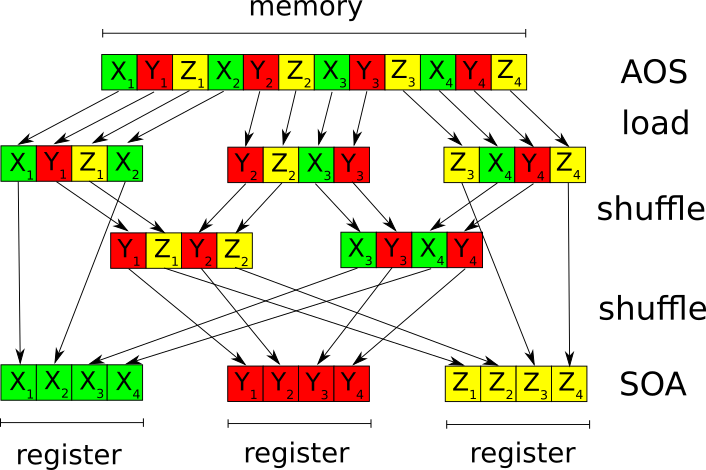}
	\end{center}
	\caption{This illustration shows how data stored in the Structures of Arrays (SoA) and Arrays of Structures (AoS) formats is loaded into registers. The parameters X, Y, and Z are part of their respective data sets T$_1$, T$_2$, T$_3$, and T$_4$. A CPU core with AVX technology uses registers to process 4 parameters simultaneously, but this requires a homogeneous memory layout. Data stored in the SoA format provides this homogeneous memory layout without any additional operation and can be processed after being loaded into the registers. Data stored in the AoS format is first loaded into the registers and subsequently rearranged by shuffling the data between the registers. These shuffle operations consume time and thus lead to lower performance.}\label{figure_SOA_register}
\end{figure*}

We implement a performance-optimized CPU version of the gradient computation in \texttt{C++}\footnote{\texttt{C++} is a programming language standardized by the International Organization for Standardization, public website: \url{https://isocpp.org/}} by using the following techniques. First, we structure our data in the Structures of Arrays (SoA) format instead of the Arrays of Structures (AoS) format, see figure~\ref{figure_SOA}. To illustrate the difference, we take a look at the internal representation of five SIS potentials using SoA and AoS. In the AoS format, they are stored as an array comprised of five different data structures. Each data structure corresponds to a SIS potential and it contains the respective data of the SIS like $\theta_{\text{center},1}$, $\theta_{\text{center},2}$, and $\sigma_{v}$. In the SoA format, the potentials are stored in one data structure which consists of different arrays. Each array corresponds to a SIS parameter like $\theta_{\text{center},1}$ and array element number one of the $\theta_{\text{center},1}$-array would correspond to the $\theta_{\text{center},1}$ location of SIS number one, element number two to the $\theta_{\text{center},1}$ location of SIS number two etc. As a result, the SIS parameters occupy contiguous parts of the memory, which is usually beneficial for vectorized computations \citep[e.g.,][]{Eijkhout2016,Besl2013}. Second, we use Advanced Vector Extensions (AVX) technology available on the latest CPU generations to harvest their built-in vectorization potential. For example, \citet{Besl2013} obtained a significant speed-up by using SoA and AVX. CPU cores with AVX technology use registers with a width of 256 bits to process 8 single precision or 4 double precision values simultaneously, see figure~\ref{figure_SOA_register} \citep{Besl2013}. Note that the same operation has to be performed for each of the simultaneously processed data values. Therefore AVX is a Single Instruction Multiple Data (SIMD) parallelism technique \citep{Eijkhout2016}. Third, we parallelize the computation using Open Multi-Processing (OpenMP)\footnote{OpenMP is an application programming interface managed by the non-profit OpenMP Architecture Review Board, public website: \url{http://www.openmp.org}} on the outermost loop of algorithm~\ref{algorithm:gradient_computation}. Each core of the multi-core CPU will now work on computing the total gradient for its assigned pixel and thus we compute the gradients for several pixels in parallel.\\
\\
We implement the GPU version of the algorithm with CUDA\footnote{CUDA is a parallel computing platform and programming model for general computing on GPUs managed by Nvidia Corporation, public website: \url{https://developer.nvidia.com/cuda-zone}}. First, we structure our data again in the SoA format. Second, we use the massively parallel architecture of GPUs to parallelize the gradient computation. Modern GPUs have many Streaming Multiprocessors (SM), which in turn consist of many Streaming Processors (SP), so the total amount of processor cores is computed by multiplying the two \citep[e.g.,][]{Eijkhout2016}. The number of cores available depends on the GPU model, for example the Nvidia Tesla P100 (henceforth called P100) possesses 3584 cores for single precision computations \citep{NvidiaCorporation2016}. In addition, GPUs are designed to be extremely efficient at switching between threads, where all threads in a single block of threads execute the same instruction \citep{Eijkhout2016}. Therefore we can effectively use many more threads than we have GPU cores. Different blocks of threads can be processed independently. This GPU parallelism is called Single Instruction Multiple Thread (SIMT) \citep{Eijkhout2016}. We use GPU threads to parallelize the outermost loop of algorithm~\ref{algorithm:gradient_computation}. Each GPU thread computes the total gradient for its assigned pixel. Therefore we can compute the gradients for thousands of pixels simultaneously.

\section{Finite machine precision errors in strong lensing}\label{section_finite_precision_errors}

\subsection{Single and double precision}\label{subsection_single_double_precision}
Modern computers usually store real numbers in the IEEE 754 single precision floating-point representation (henceforth called SP) or the IEEE 754 double precision floating-point representation (henceforth called DP) \citep[][see e.g. \citet{Goldberg1991} for an overview of floating-point arithmetic]{IEEE2008}. A real number $x \in \mathbb{R}$ in decimal representation is thus stored in a binary format, 
\begin{equation}
	x = \sigma \times \bar{x}_{2} \times 2^{e},
\end{equation} 
where the integer $e$ is the exponent, the sign $\sigma$ equals $+1$ or $-1$, and $\bar{x}_{2}$ is a binary number satisfying $(1)_{2} \leq \bar{x}_{2} < (10)_{2}$ \citep{IEEE2008}. Note that the binary number $\bar{x}_{2}$ consists of several integer digits $d \in \{0,1\}$, i.e. $\bar{x}_{2} = d_{0}.d_{1} d_{2} \dots d_{p-1}$. In the remainder of this paper we will denote the binary format by using the subscript $2$, so $(1)_{2}$ and $(10)_{2}$ correspond to the numbers $1$ and $2$ in decimal representation. For example, the number 2.25 would correspond to $\sigma = +1$, $\bar{x}_{2} = (1.001)_{2}$, and $e = (1)_{2}$\footnote{In practice, the leading bit of $\bar{x}_{2}$ would be implicit and $e$ would be stored as a biased exponent, but we can ignore such intricacies here to simplify the presentation}. The number of digits in $\bar{x}_{2}$ is called the precision $p$ of the representation. According to IEEE 754, SP has a precision of $p = 24$ digits and an exponent $-126 \leq e \leq 127$, while DP has $p = 53$ and $-1022 \leq e \leq 1023$. SP values are stored using 4 bytes (= 32 bits) and DP values using 8 bytes (= 64 bits) \citep{IEEE2008}. As a result, DP can store a number $x$ with higher accuracy than SP, but this comes at the price of increased memory consumption and usually also reduced computing performance \citep[e.g.,][]{Besl2013,Eijkhout2016} \\
\\
Both DP and SP have only a limited amount of memory available and thus their accuracy is limited. We define the machine epsilon $\epsilon$ as the difference between 1 and the next larger number that can be stored using the given representation \citep{Eijkhout2016}. For SP and DP we thus have respectively $\epsilon = 2^{-23} \approx 1.2 \times 10^{-7}$ and $\epsilon = 2^{-52} \approx 2.2 \times 10^{-16}$. These errors are so small that they might seem unimportant at first, but they will be magnified by the different computing operations performed in the course of an algorithm, so that they can become very large and relevant once the final result is obtained. \\
\\
To illustrate this point, we now look at a hypothetical calculator\footnote{This illustration is inspired by an example in the lecture notes of Catalin Trenchea, available online at \url{http://www.math.pitt.edu/~trenchea/math1070/MATH1070_2_Error_and_Computer_Arithmetic.pdf}}. For simplicity, it does not use SP or DP, but a decimal number representation with 6 digits precision and no exponent. We compute a relatively simple function, $f(x) = x \times (\sqrt{x+1} - \sqrt{x})$. For $x = 50,000$, the result from the hypothetical calculator is 100, while the true result is 111.8, so we have a relative error of more than 10\%. To understand this behavior, we take a look at the different steps which the calculator has to perform. It computes $\sqrt{50,001}$ and rounds the result to 6 digits (result: 223.609) and then it repeats these steps for $\sqrt{50,000}$ (result: 223.607). Therefore we have two rounding errors, but they are very small. However, now the calculator subtracts two almost equal numbers to obtain $000.002$, so only the last number of the result is a significant digit. We have lost a lot of accuracy which we cannot recover. The subsequent multiplication does not increase the error, but it propagates it into the final result. This example illustrates that even with the high precision available in modern computers, the result of a sufficiently long and complex algorithm can be significantly affected by the chosen number representation. \\
\\
DP permits a much higher accuracy than SP and therefore it is tempting to simply use it for all computations. However, this accuracy comes at the price of computing performance. As shown in table~\ref{table_theoretical_GFLOPS}, this is particularly true for GPUs. While the theoretical maximum computing performance of a modern CPU decreases by a factor of two, the peak performance of a consumer GPU like the Nvidia GeForce GTX 1080 Ti (henceforth called GTX) drops by two orders of magnitude. This is a significant problem for GPU-accelerated scientific software, where SP is often not accurate enough. To ameliorate this issue, graphics card manufacturers introduced new hardware specifically designed to improve the DP performance. The P100 and its recently released successor, the Nvidia Tesla V100 (henceforth called V100) achieve half of their SP performance when using DP. However, these special purpose GPUs are much more expensive than regular consumer GPUs like the GTX, typically by an order of magnitude. Table~\ref{table_theoretical_GFLOPS} shows that the SP performance of a high-end consumer GPU is comparable to the SP power of the special purpose GPUs. Thus, if it is possible to use SP instead of DP in our lensing algorithm, we would not only significantly increase the code performance on both CPUs and GPUs, but we might also be able to achieve a close to optimal performance with relatively cheap hardware.

\begin{center}
\begin{table*}
\centering
	\begin{tabular}{c|c|c|c|c}
	    &Intel Xeon E5-2680 v3 & Nvidia GTX 1080 Ti & Nvidia P100 & Nvidia V100\\
	    &12 cores, 2.50~GHz & 3584 cores, 1582~MHz  & 3584 cores, 1480~MHz & 5120 cores, 1530~MHz\\
	    \hline
	    Double precision & 240~GFLOPS & 354~GFLOPS & 5304~GFLOPS & 7833~GFLOPS\\
	    Single precision & 480~GFLOPS & 11340~GFLOPS & 10609~GFLOPS & 15667~GFLOPS
	\end{tabular}
\caption{Theoretical maximum computing performance for our used CPU and GPU models. These values can only serve as a rough indicator of expected performance, as the real application performance will depend on many parameters such as the used algorithm and its implementation. We list the base frequency for the CPU while we use the boost frequency for the GPUs, as the CPU typically reaches the boost frequency only on a few cores and and not on all cores simultaneously.  We compute the CPU maximum computing performance using the following formula: Two operations per cycle $\times$ frequency $\times$ AVX vectorization $\times$ number of cores \citep{Besl2013}. Note that the AVX factor for SP is two times larger than for DP. We use the same formula, but without the AVX factor, for GPUs. Graphics cards have a different number of cores for SP and DP computations and thus a different maximum performance. The number of GPU cores listed in the table is for SP computations. Due to the lower number of DP cores, the Nvidia GTX 1080 Ti's GP102 GPU has thirty-two times less performance in DP computations than in SP \citep[e.g.,][]{Harris2016}, while the P100's GP100 GPU and the V100's GV100 GPU are two times slower \citep{NvidiaCorporation2016,NvidiaCorporation2017a}. The number of cores and the frequencies are taken from \citet{IntelCorporation2014} and \citet{,NvidiaCorporation2017,NvidiaCorporation2017a,NvidiaCorporation2016}.} \label{table_theoretical_GFLOPS}
\end{table*}
\end{center}

\subsection{Computing finite precision errors for strong lensing} 
We will now show that using SP in algorithm~\ref{algorithm:gradient_computation} is accurate enough for a large fraction of the image pixels. We restrict ourselves again to the SIE model. It is possible to generalize these results to other parametric models, but the fraction of the image for which SP is accurate enough will vary and has to be computed for each model independently.\\
\\
The lens equation~\ref{equation:lens_equation_reformulated} maps the triangular pixels in the image plane onto triangular pixels in the source plane. We assume a HFF pixel size of 0.03~arcsec and we maximize the lensing effect by using $D_{\text{LS}}/D_{\text{OS}}= 1$. As a result, the lens equation is now a simple subtraction of $\nabla \Psi_{\epsilon}(\theta_{1},\theta_{2})$. We now look at an observed multiple image in the image plane. Note that our ability to locate the multiple image is observationally constrained by the size of the image pixels, so there is an observational error budget on the image location of half a pixel. In addition, the algorithm lenses both the triangular pixel and the image into the source plane. It is possible that their respective errors due to machine precision have the same magnitude but the opposite sign, and therefore the error budget shrinks by another factor of two. As a result, the value of $\nabla \Psi_{\epsilon}$ can be considered accurate enough if the error $E$ is smaller than a quarter of a pixel. Thus our upper limit for the gradient error is $E_{i} \leq 7.5 \times 10^{-3}~\text{arcsec}$, where $i = 1,2$.\\
\\
However, this error budget does not yet account for the magnification effect of strong lensing. Background sources and distance scales appear magnified when they are strongly lensed and consequently distance scales in the image plane like pixel sizes will be be de-magnified when they are mapped into the source plane. The resulting error budget for $\nabla \Psi_{\epsilon}$ becomes thus $E_{i} \leq 7.5 \times 10^{-3}~\text{arcsec}/M_{i}$, where $M_{i}$ is the magnification along the $\theta_{i}$-axis.  \\
\\
 In \ref{appendix_a} we derive an upper bound for the error of $\nabla \Psi_{\epsilon}$ due to finite machine precision. We assume that the lens is a strong lensing cluster modeled with two cluster-scale SIE halos. The SP upper error bound along the $\theta_{i}$-axis is $\Delta (\nabla \Psi_{\epsilon}) \leq 2.3 \times 10^{-3}~\text{arcsec}$ if we use the following approach. As discussed in \ref{appendix_a}, we compute the gradients with SP except in pixel grids of $400 \times 400~\text{pixels}$ around cluster-scale halos and $20 \times 20~\text{pixels}$ around galaxy-scale halos, where we use DP. This corresponds to approximately 1\% of all image pixels. As a result, SP is accurate enough for each of the remaining 99\% of the image pixels if the respective magnification along both $\theta_{i}$-axes is $M_i \leq 3.26$. In strong lensing, we typically measure the magnification of the area of a multiple image and not the magnification along an axis. The measured values are typically single digits \citep[see e.g.][for magnification values of a HFF cluster]{Jauzac2015b}. While these values cannot easily be converted to axis-magnifications due to the the typically arc-like shape of strongly magnified images, they strongly suggest that SP will be accurate enough for a large fraction of the image. However, strong lensing clusters have critical lines where the magnification diverges. In the case of the SIS, this critical line is the Einstein ring. While the magnification does not become infinite in practice \citep[see e.g.][for a detailed discussion]{Bartelmann2001}, it can become very large and thus SP will no longer be accurate enough. Consequently, we can use SP for a large fraction of the image, but we also need to implement a mechanism which ensures that we compute the gradients with DP whenever SP is not enough due to high magnification. 

\newpage

\subsection{Fixing the missing accuracy close to critical lines} 
\begin{figure}
	\begin{center}
	    \includegraphics[width=0.4\textwidth]{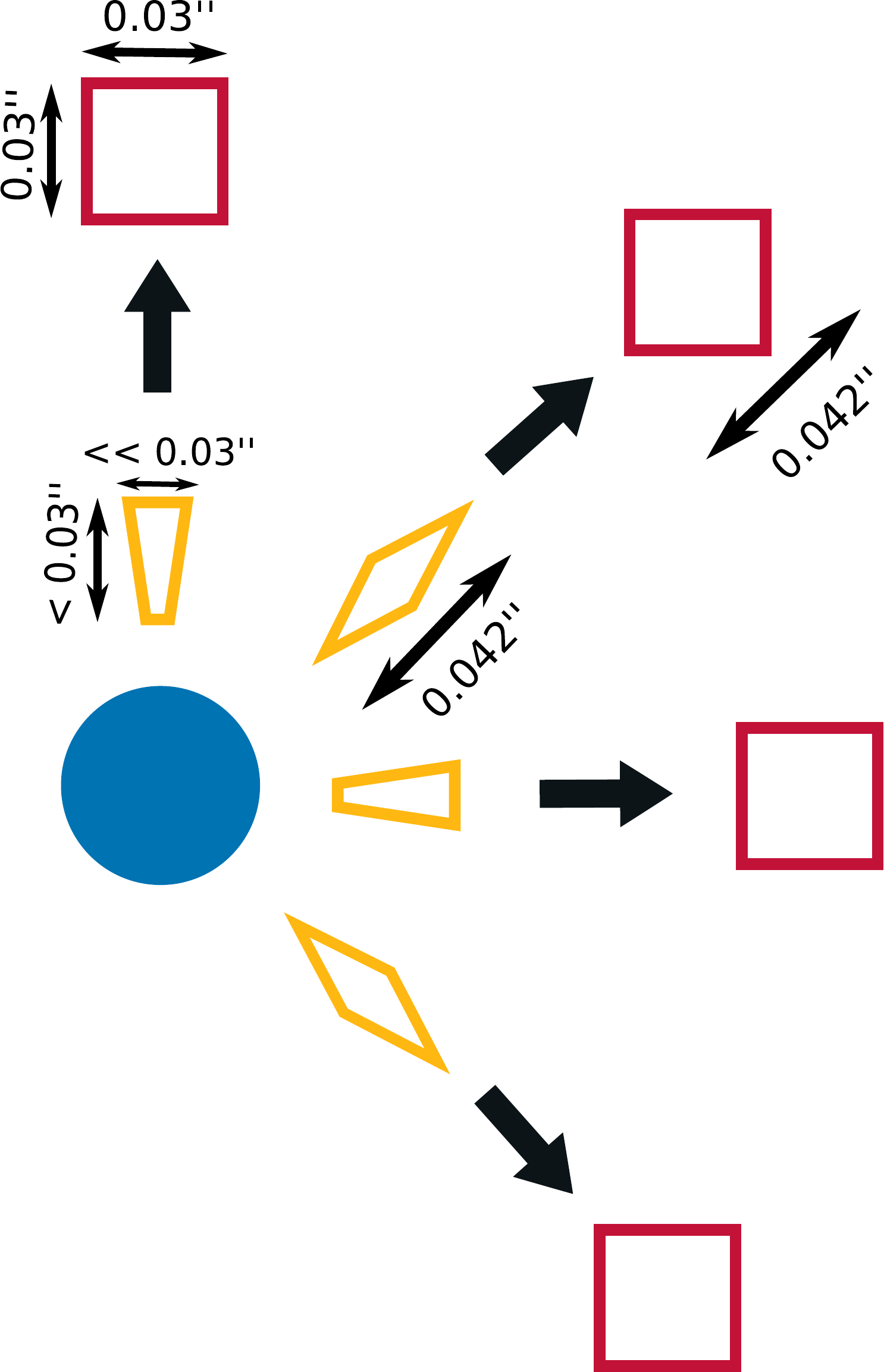}
	\end{center}
	\caption{The source plane pixels (yellow) are greatly distorted with respect to the corresponding regular image plane pixels (red). This example shows the distortions caused by a single symmetrical SIS lens (blue) for angles of -45, 0, 45, and 90~degrees. Note that the greatest distortion occurs perpendicular to the lensing direction, but small lensing effects can also occur alongside this direction, as the example for the 90~degree angle shows. The magnitude of this effect is typically negligible compared to the perpendicular distortion. }\label{figure_distortions_angles}
\end{figure}

\begin{figure}
	\begin{center}
		\includegraphics[width=0.2\textwidth]{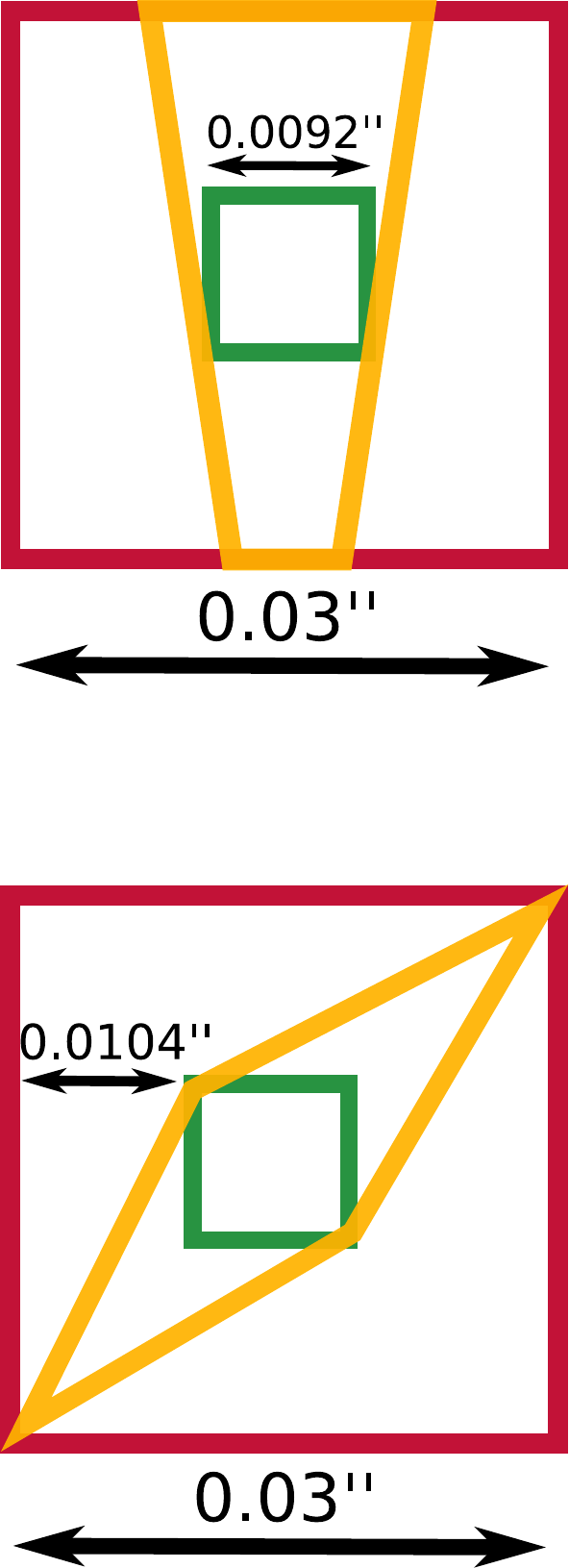}
	\end{center}
	\caption{The values of $\nabla \Psi_{\epsilon}$ must be recomputed in DP if the distorted and de-magnified source plane pixels (yellow) become smaller than the error due to finite machine precision (green box). The regular image plane pixel is overplotted in red. The top figure illustrates the lensing example with an angle of 90~degrees shown in figure~\ref{figure_distortions_angles} and the bottom figure demonstrates the example with an angle of 45~degrees. }\label{figure_distortion_criterions}
\end{figure}

\begin{figure*}
	\begin{center}
		\includegraphics[width=\textwidth, trim= 0cm 8cm 0cm 8cm, clip]{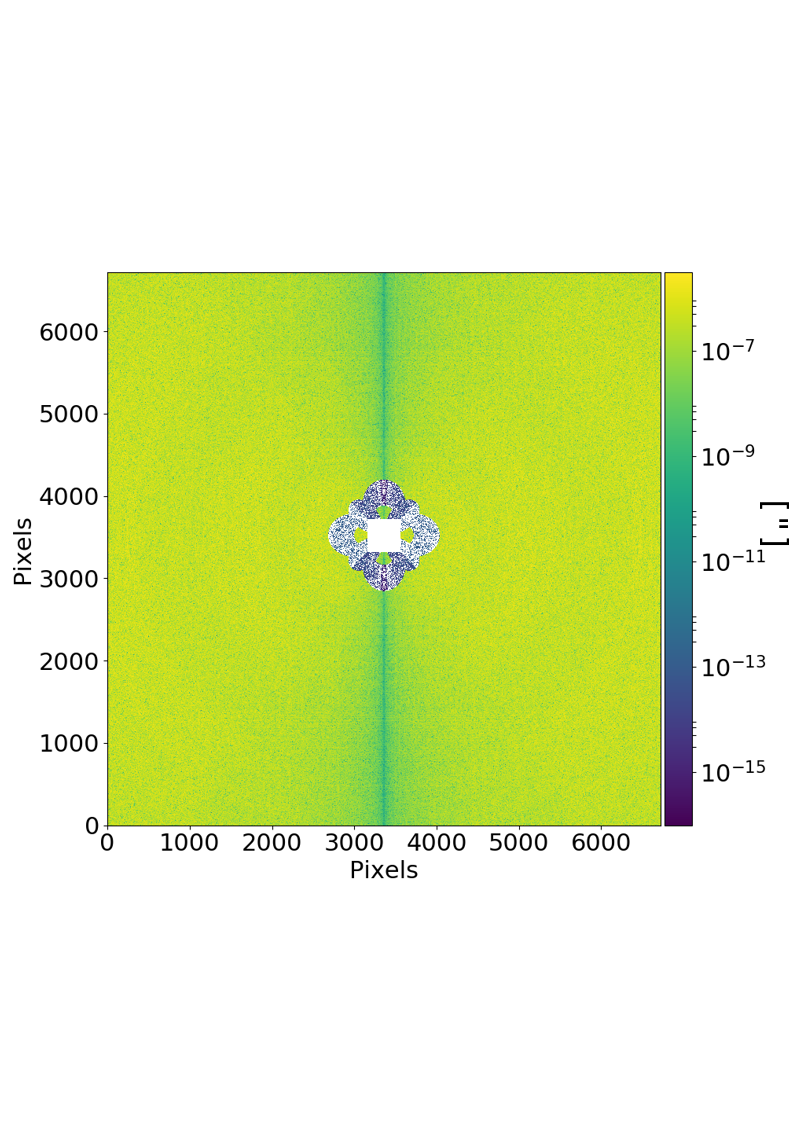}
	\end{center}
	\caption{Difference between the values of $(\nabla \Psi_{\epsilon})_{1}$ computed with our mixed precision and DP algorithms for a single spherical cluster-scale SIS lens. The color white indicates a difference of zero. The green pixels are calculated with SP and the white and blue pixels are re-computed with DP. The error for every pixel is within the allowed error bounds. The rectangular patch around the lens center in which we use DP is clearly visible.   }\label{figure_Difference_Mixed_Double_OneSIS}
\end{figure*}

We add the missing accuracy close to critical lines as follows. First, we compute $\nabla \Psi_{\epsilon}$ for each pixel in the image using the approach presented in the previous subsection. Second, we compute for each pixel
\begin{align}
	\delta_{1}(\theta_{1},\theta_{2})  &= (\nabla \Psi_{\epsilon})_{1}(\theta_{1},\theta_{2}) - (\nabla \Psi_{\epsilon})_{1}(\theta_{1}-\Delta x,\theta_{2}), \nonumber \\
	\delta_{2}(\theta_{1},\theta_{2})  &= (\nabla \Psi_{\epsilon})_{2}(\theta_{1},\theta_{2}) - (\nabla \Psi_{\epsilon})_{2}(\theta_{1},\theta_{2}-\Delta x),
\end{align}
which is computationally cheap because we have already computed the gradient values for all pixels. For the HST ACS, we have a pixel height and width $\Delta x = 0.03~\text{arcsec}$. Note that $\delta_{i}$ corresponds to the change of the pixel length along the $\theta_{i}$-axis due to lensing into the source plane. Third, we recompute $\nabla \Psi_{\epsilon}$ in DP for all pixels where 
\begin{align}
	|0.03~\text{arcsec} - \delta_{i}(\theta_{1},\theta_{2})| < 0.0092~\text{arcsec}\label{equation:Criterion_delta_1_2}
\end{align}
for $i=1$ or $i=2$, which implies that $M_{i} > 3.26$. We derive this condition by computing the pixel length in the source plane $\Delta x_{\text{source}}$ along the $\beta_{1}$-axis,
\begin{align}
	|\Delta x_{\text{source},1}(\theta_{1},\theta_{2})|  &= |\beta_{1}(\theta_{1},\theta_{2}) - \beta_{1}(\theta_{1}-\Delta x,\theta_{2})|  \nonumber \\
	&= |\Delta x  - \delta_{1}(\theta_{1},\theta_{2})|,
\end{align}
where we used the lens equation. An analogous relation holds for the $\beta_{2}$-axis. Note that taking the absolute value of $\Delta x_{\text{source}}$ is necessary because lensing can change the image parity \citep[see e.g. the review][]{Kneib2011}.  For the assumed HFF pixel scale, the condition that $M_{i} > 3.26$ translates into $\Delta x_{\text{source},i} < 0.0092~\text{arcsec}$. The lensing effect along the $\theta_{1}$- and $\theta_{2}$-axis is shown in figure~\ref{figure_distortions_angles} and the criterion in equation~\ref{equation:Criterion_delta_1_2} is illustrated in the top part of figure~\ref{figure_distortion_criterions}.\\
\\
We can assume that each source is lensed along a chosen $\theta_{i}$-axis, as this can be achieved by a simple change of the image plane coordinate system. However, the shape of the image plane pixels is not invariant under such a transformation, as figure~\ref{figure_distortions_angles} illustrates. Therefore we need two additional criteria. We define   
\begin{align}
\delta_{3}(\theta_{1},\theta_{2})  &= (\nabla \Psi_{\epsilon})_{1}(\theta_{1},\theta_{2}) - (\nabla \Psi_{\epsilon})_{1}(\theta_{1},\theta_{2}-\Delta x), \nonumber \\
\delta_{4}(\theta_{1},\theta_{2})  &= (\nabla \Psi_{\epsilon})_{2}(\theta_{1},\theta_{2}) - (\nabla \Psi_{\epsilon})_{2}(\theta_{1}-\Delta x,\theta_{2}),
\end{align}
and we recompute $\nabla \Psi_{\epsilon}$ in DP if
\begin{align}
| \delta_{i}(\theta_{1},\theta_{2}) | > 0.0104~\text{arcsec} \label{equation:Criterion_delta_3_4}
\end{align}
for $i=3$ or $i=4$. This case is illustrated in the bottom part of figure~\ref{figure_distortion_criterions}. \\
\\
In summary, we compute $\nabla \Psi_{\epsilon}$ in SP everywhere except in small patches centered on the origin of each lens as described in \ref{appendix_a} and for the pixels where the criteria defined in equations~\ref{equation:Criterion_delta_1_2} and \ref{equation:Criterion_delta_3_4} hold. This is illustrated in figure~\ref{figure_Difference_Mixed_Double_OneSIS}.

\section{Performance measurements}\label{section_benchmarks}
\begin{figure*}
	\begin{center}
		\includegraphics[width=\textwidth, trim= 0cm 8cm 0cm 8cm, clip]{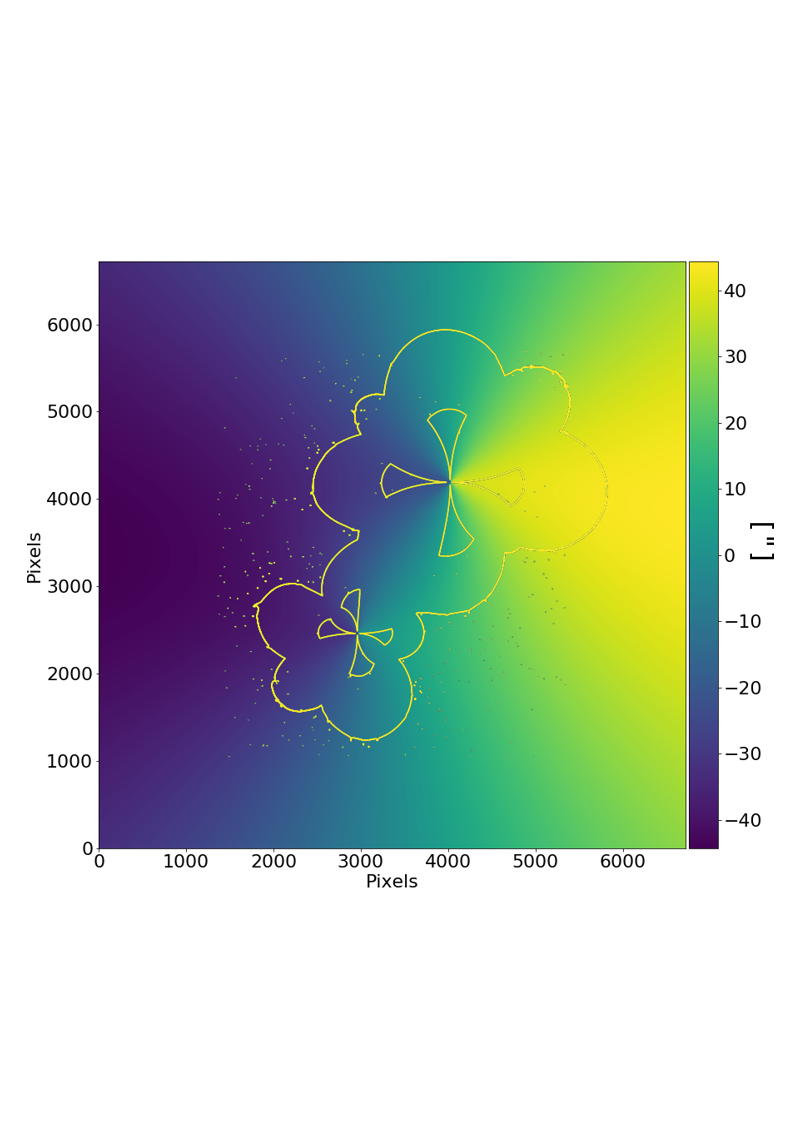}
	\end{center}
	\caption{Values of $(\nabla \Psi_{\epsilon})_{1}$ computed in DP for the HFF-like galaxy cluster lens used for the performance benchmark. The yellow contours indicate the area in which at least one of the conditions shown in equations~\ref{equation:Criterion_delta_1_2} and~\ref{equation:Criterion_delta_3_4} is triggered. The patches for the halo centers are not displayed. }\label{figure_gradient_x_benchmark}
\end{figure*}

\begin{figure*}
	\begin{center}
		\includegraphics[width=\textwidth, trim= 0cm 7cm 0cm 10cm, clip]{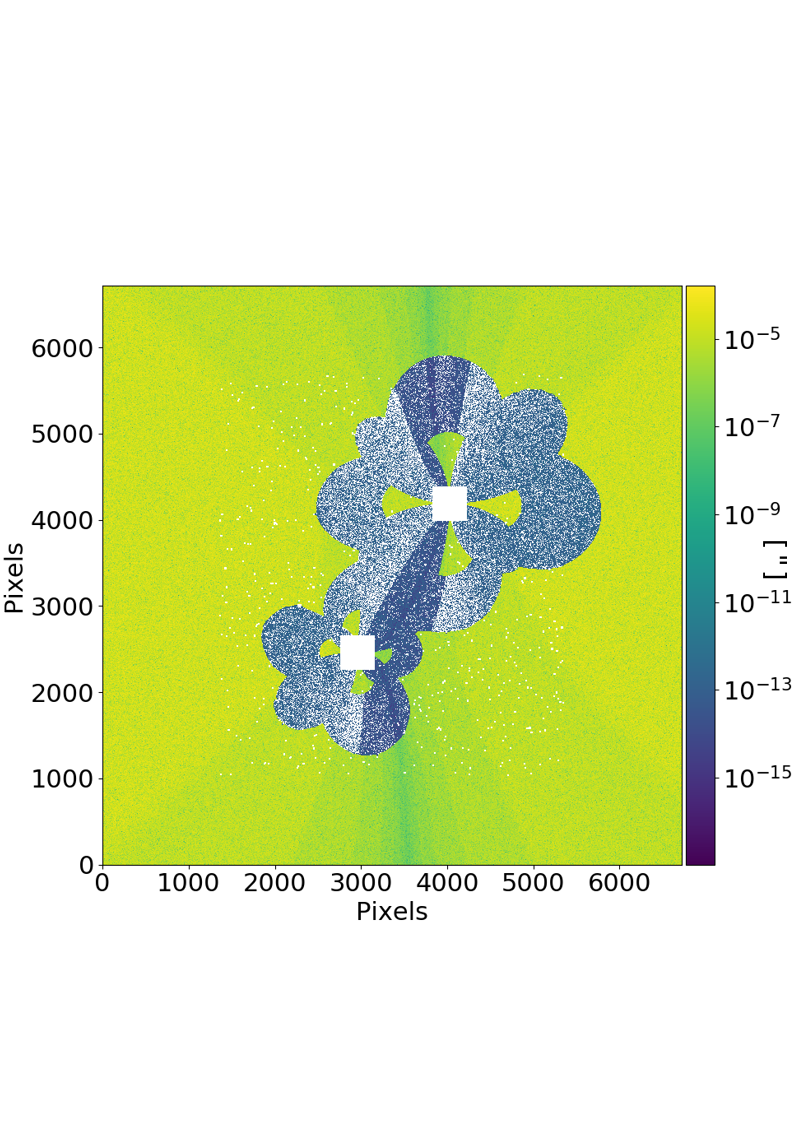}
	\end{center}
	\caption{Difference between the values of $(\nabla \Psi_{\epsilon})_{1}$ computed with our mixed precision and DP algorithms for the HFF-like galaxy cluster lens used for the performance benchmark. The color white indicates a difference of zero. The green pixels are calculated with SP and the white and blue pixels are re-computed with DP. The error for every pixel is within the allowed error bounds. The small rectangular patches around the 700 galaxy halo centers in which we use DP are clearly visible.   }\label{figure_Difference_Mixed_Double_benchmark}
\end{figure*}

\begin{figure}
	\begin{center}
		\includegraphics[width=0.45\textwidth, trim= 0cm 7cm 0cm 11cm, clip]{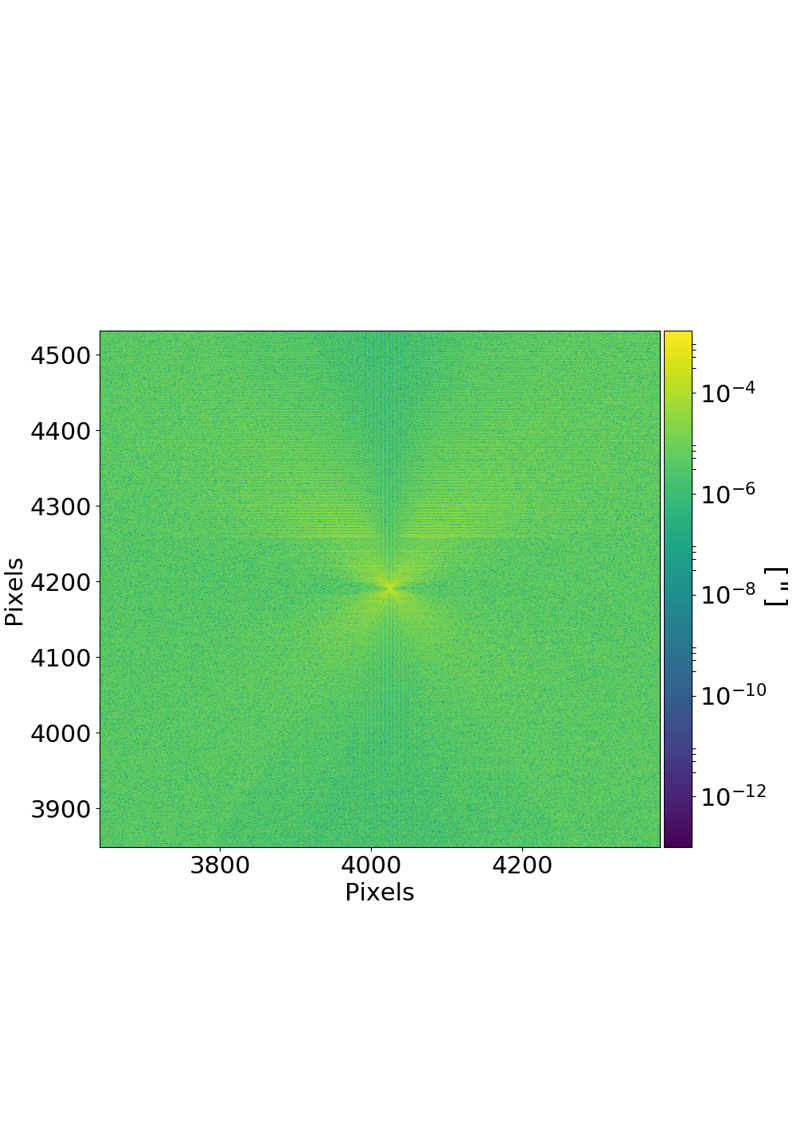}
	\end{center}
	\caption{Difference between the values of $(\nabla \Psi_{\epsilon})_{1}$ computed with a pure SP algorithm and a DP algorithm. The figure shows a zoom-in on a cluster-scale SIE halo of the HFF-like galaxy cluster used in the performance benchmark. The areas with the largest errors follow clearly the pattern shown in figure~\ref{figure_Difference_Mixed_Double_OneSIS}. Therefore our mixed precision algorithm would re-compute these pixels with DP and thus ensure the accuracy of the result.   }\label{figure_Difference_Single_Double}
\end{figure}

We implement both the GPU and the CPU version of the gradient computation twice, once in SP and once in DP. The respective versions are identical up to the change in precision. In addition, we implement the mixed precision algorithm for both types of hardware. In the first step, this algorithm computes the result for each pixel in SP. In the second step, it checks which results are not accurate enough and recomputes these with DP. For this purpose, the algorithm uses the criterions developed in the previous section. The GPU implementation of the mixed precision algorithm uses asynchronous computations and load balancing for the second step, i.e. the computation of the most expensive DP calculations can be dispatched asynchronously between the CPU and the GPU. We adjust the load balancing for the different GPU models. As table~\ref{table_theoretical_GFLOPS} shows, this is particularly interesting for the GTX: The use of a hybrid CPU/GPU approach alleviates the very low DP performance of this card and drastically reduces the impact of the DP computations on the overall run time.\\
\\
In the next step, we want to measure the performance gain of using HPC methods in strong lensing. For this purpose, we measure the time which the different software implementations require to compute the gradient of a HFF-like cluster lens for each pixel of a \textit{Hubble} image. We have repeated this measurement several times and find that the benchmark results are stable, i.e. they do not significantly vary in different runs with the same setup. We also compute the gradients with the current \texttt{LENSTOOL} software, which serves as a reference. We assume a $\Lambda$CDM cosmology with $H_{0} = 70~\text{km/(s~Mpc)}$, $\Omega_{m} = 0.3$, and $\Omega_{\Lambda} = 0.7$. We use an image with $6730 \times 6730~\text{pixels}$ and a pixel scale of 0.03~arcsec/pixel to simulate images from the HST ACS. The galaxy cluster consists of two cluster-scale and 700 galaxy-scale halos like in the HFF cluster Abell~2744 \citep[e.g.,][]{Jauzac2015b}.  We model the lens using SIE halos. The lens redshift is 0.3 and all sources are at the same redshift $z_{\text{source}} = 2.0$. The velocity dispersion determined by \citet{Jauzac2015b} for one cluster-scale halo of Abell~2744 is approximately 1200~km/s and for a galaxy halo it is roughly 150~km/s. However, these values correspond to the velocity dispersion parameter of the parametric lens model chosen in \citet{Jauzac2015b}, which is not identical to measured line-of-sight velocity dispersions of galaxies. The exact conversion must be computed numerically, but for our purposes a rough agreement is enough, so we can use a conversion factor of 0.85 \citep{Eliasdottir2007}. This leads to $\sigma_{v} \approx 1000~\text{km/s}$ and we choose this value and a pseudo-ellipticity of the potential $\epsilon = 0.05$ for the first cluster scale halo. For the second large-scale halo, we use $\sigma_{v} = 700~\text{km/s}$ and $\epsilon = 0.04$. We model the galaxy halos by allowing $\sigma_{v}$ and $\epsilon$ to randomly vary between 10 and $15~\text{km/s}$ and 0 and 0.15, respectively. Note that the galaxy halo velocity dispersions are approximately a factor ten smaller than the ones used in \citet{Jauzac2015b}, because the magnitude of the scaled deflection angle for a SIE does not decrease with distance from the lens center, as it does for more realistic parametric lens models. Thus we need to decrease the velocity dispersion to limit the lensing effect of individual galaxies at large separations from the galaxy.\\
\\
Figure~\ref{figure_gradient_x_benchmark} presents the gradient values in $\theta_{1}$ direction for the cluster lens. Figure~\ref{figure_Difference_Mixed_Double_benchmark} shows that the error resulting from our mixed precision algorithm is within the allowed limit for each pixel. Figure~\ref{figure_Difference_Single_Double} illustrates the hypothetical error of a pure SP algorithm close to a cluster-scale halo. We see that the area in which we re-compute $\nabla \Psi_{\epsilon}$ shown in figure~\ref{figure_Difference_Mixed_Double_OneSIS} covers nicely the area in which the SP error is largest.  \\

\begin{figure*}
	\begin{center}
		\includegraphics[width=\textwidth]{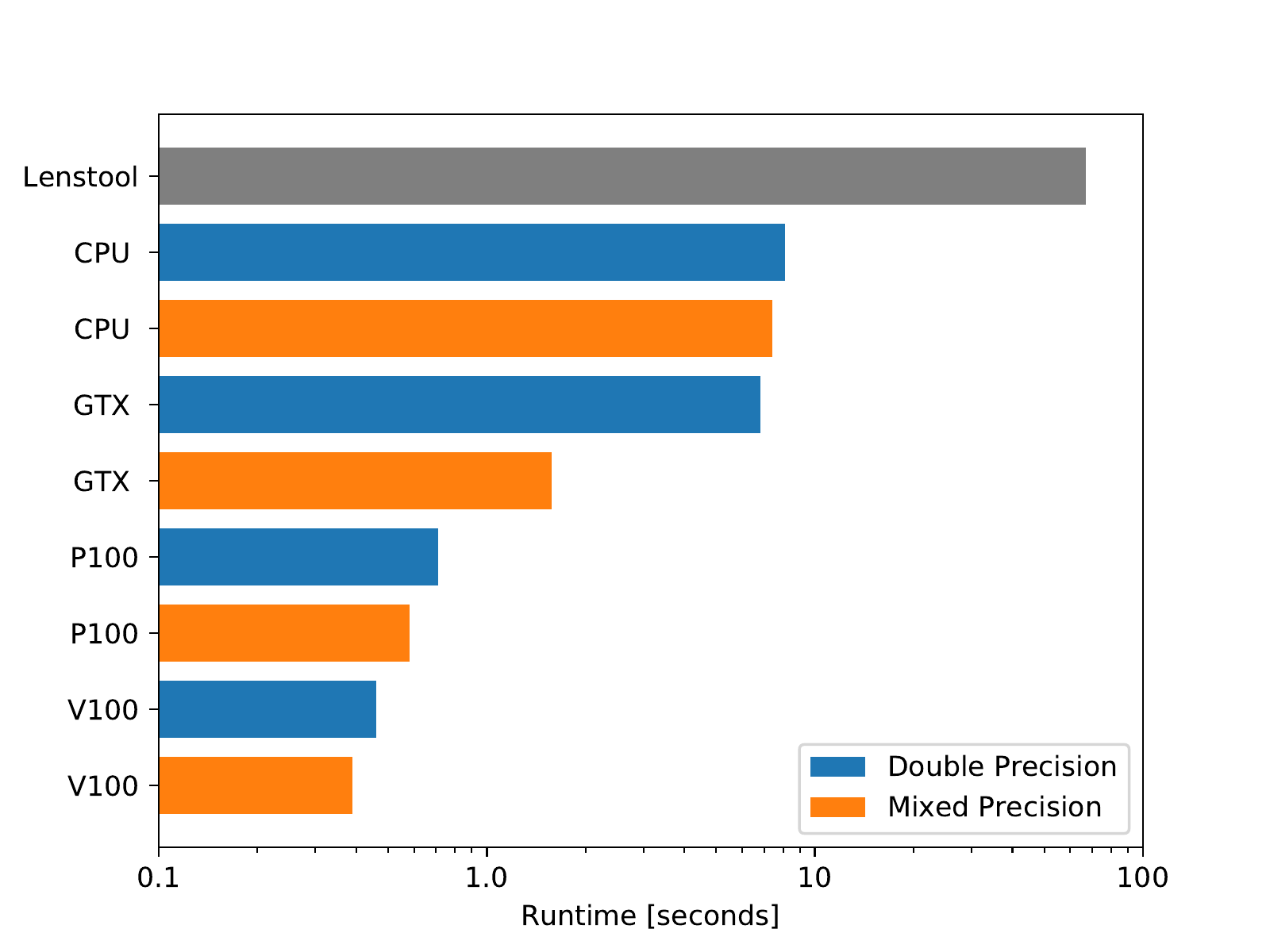}
	\end{center}
	\caption{Logarithmic plot of the benchmark performance for the different gradient computation implementations. We calculate the gradient for each pixel in the simulated HST image of a HFF-like galaxy cluster lens. The current \texttt{LENSTOOL} software serves as performance reference. The HPC-optimized implementation on the same Intel CPU with 12 cores is called CPU. It is already an order of magnitude faster. The GPU implementations can reduce the run time by another order of magnitude. Note that the mixed precision GPU algorithm uses a hybrid CPU/GPU approach. The mixed precision algorithm is faster than the DP implementation for each of the different hardware devices. }\label{figure_benchmark_results}
\end{figure*}

\begin{figure}
	\begin{center}
		\includegraphics[width=0.5\textwidth]{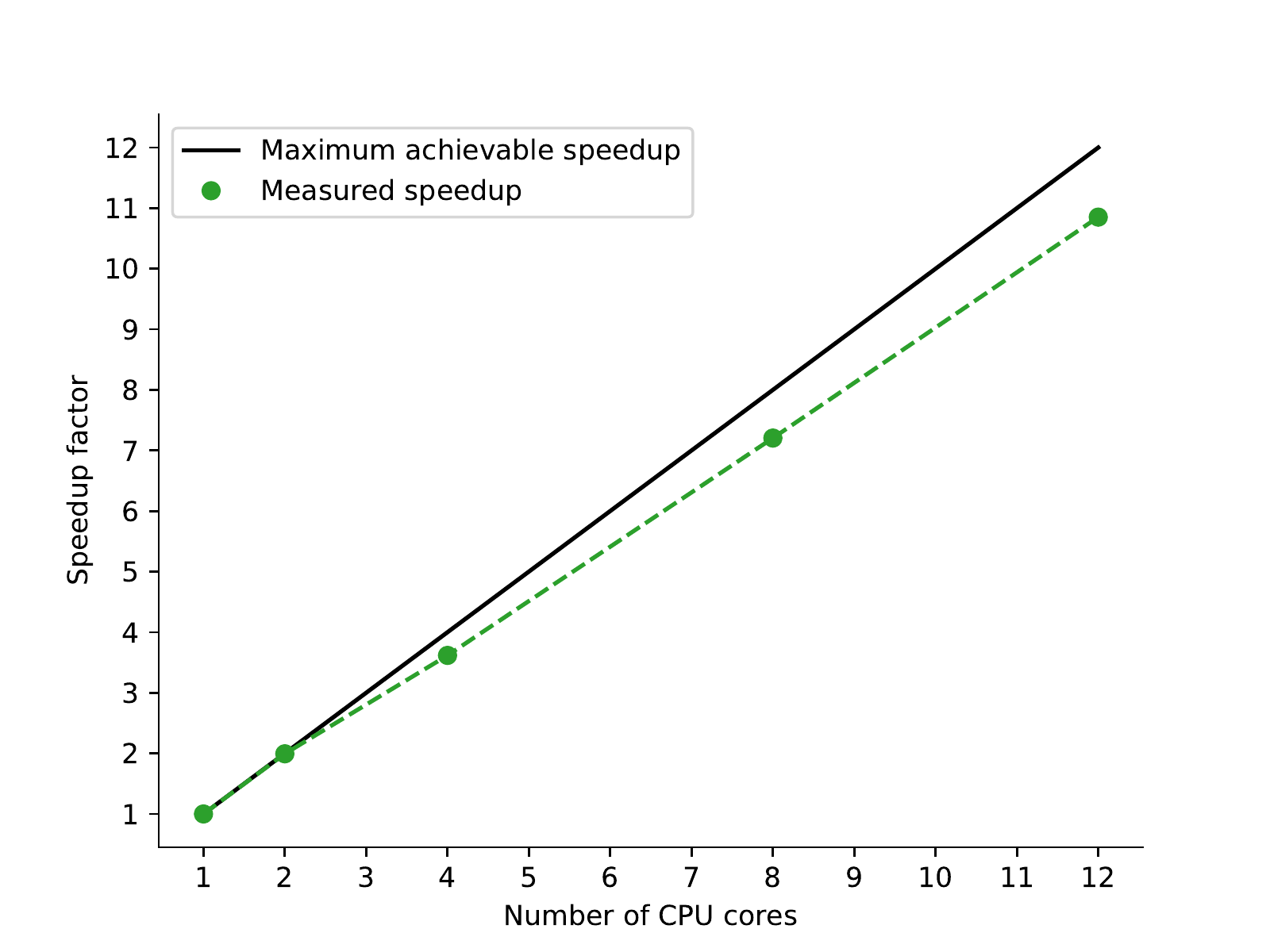}
	\end{center}
	\caption{The performance of the HPC-optimized CPU code scales almost perfectly with the number of used CPU cores on a single computing cluster node. }\label{figure_CPU_scaling}
\end{figure}

\begin{figure}
	\begin{center}
		\includegraphics[width=0.5\textwidth]{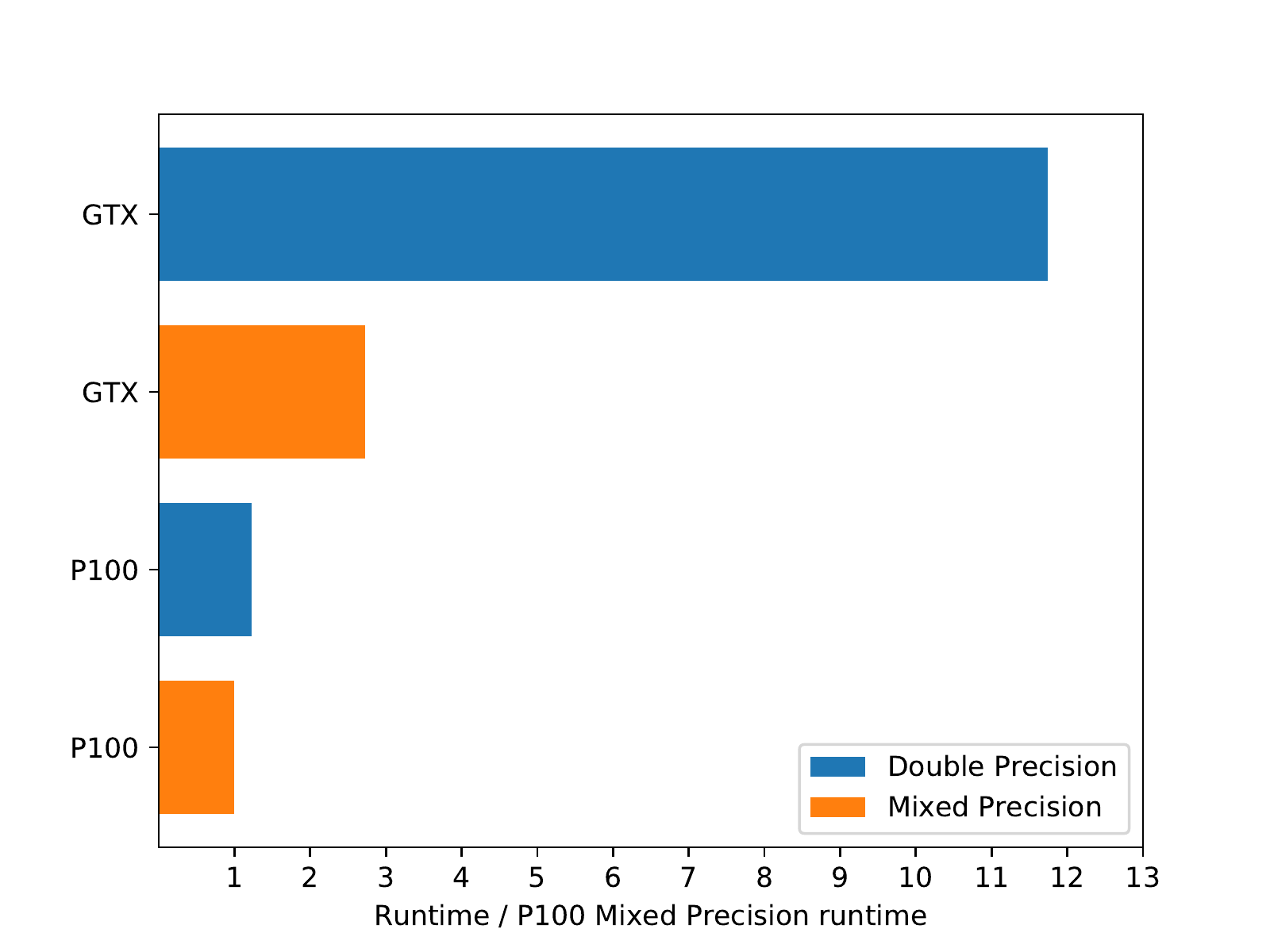}
	\end{center}
	\caption{Benchmark performance for a high-end GPU (Nvidia P100) and the consumer graphics card (Nvidia GTX). Both GPUs are based on the Pascal GPU architecture. The Nvidia V100 is based on the more recent Volta architecture and thus not shown in this comparison. The P100 is an order of magnitude more expensive than the GTX and especially designed for DP performance in scientific applications. Consequently it clearly outperforms the GTX when only DP is used. However, the mixed precision algorithm in combination with the hybrid CPU/GPU approach greatly accelerates the performance of the GTX and its run time comes close to the P100. }\label{figure_P100_vs_GTX}
\end{figure}

Table~\ref{table_benchmark} and figure~\ref{figure_benchmark_results} present the benchmark performance of the different gradient computation implementations. The speedup of the HPC-optimized codes with respect to the current \texttt{LENSTOOL} software is considerable. The indicated \texttt{LENSTOOL} performance is obtained by using all 12 CPU cores. It is thus the best currently achievable speed, as \texttt{LENSTOOL} cannot be run on multiple computer nodes and its performance is thus limited by the number of CPUs available on a single node. In addition, the mixed precision implementations are consistently faster than the DP ones, which validates our approach. The HPC CPU version reduces the run time of the benchmark by one order of magnitude and the GPU implementations by up to two while keeping the error within the allowed bounds. The consumer-grade GTX card displays the largest performance gain with respect to the DP computation. Figure~\ref{figure_CPU_scaling} demonstrates that our HPC-optimized CPU software scales almost perfectly with the number of cores available on a single node using multi-threading. Figure~\ref{figure_P100_vs_GTX} compares the benchmark performance of the Nvidia GTX with the high-end GPU based on the same Pascal GPU architecture (Nvidia P100). The P100 is an order of magnitude more expensive than the GTX. The P100 is considerably faster when only DP is used, which is expected as it was designed for this purpose. However, as soon as the mixed precision algorithm is used, the GTX reduces its run time dramatically and the performance comes close to the P100.

\begin{center}
	\begin{table*}
		\centering
		\begin{tabular}{c|c|c|c|c}
			&Run time & Run time & Run time reduction & Speedup factor \\
			& Double Precision & Mixed Precision & Double $\to$ Mixed Precision & compared to \\
			& (seconds) & (seconds) & (\%) & \texttt{LENSTOOL} \\
			\hline
			CPU & 8.1~ & 7.40 &  ~9 & ~~9 \\
			GTX &6.8~ & 1.58 &  77 &  ~42\\
			P100 & 0.71 & 0.58 & 18 & 115 \\
			V100 & 0.46 & 0.39 &  15 & 171 \\
		\end{tabular}
		\caption{Benchmark results of the gradient computation implementations for a HFF-like lens. We show the double and mixed precision run times for the HPC-optimized CPU version using 12 cores and the three GPU models. The mixed precision algorithm for the GPUs uses a hybrid CPU/GPU implementation. Column three presents the measured run time advantage of mixed precision over double precision. Note that the mixed precision algorithm requires a substantial amount of additional computations compared to the double precision algorithm, as it must check for which pixels single precision is accurate enough and for which ones the gradient must be recomputed in double precision. Despite this overhead, the mixed precision implementation is the fastest for all four hardware devices. The fourth column shows the speedup of the mixed precision implementation with respect to the best currently achievable speed of \texttt{LENSTOOL}. } \label{table_benchmark}
	\end{table*}
\end{center}

\section{Discussion}\label{section_discussion}
 The performance measurements in the last section demonstrate clearly the value of HPC methods for strong lensing software. They lead to a speedup of one to two orders of magnitude, depending on the chosen hardware. In addition, our measurements show that GPUs are perfectly suited for the massively parallel lensing calculations. As expected, the high-end GPUs have a big performance advantage in DP computations, but our mixed precision algorithm and the hybrid CPU/GPU approach bring the consumer GPU's performance very close to its more expensive siblings. Note that the use of mixed precision also benefits the benchmark performance of the the high-end GPUs and of the CPU, but not on the same scale. Mixed precision thus leads to a performance benefit regardless of used hardware while also delivering accurate results.\\
\\
Furthermore, table~\ref{table_TDP} demonstrates that the use of HPC methods dramatically reduces the energy consumption. We estimate the required energy to solution of the respective gradient computation implementations by multiplying the Thermal Design Power (TDP) of the used hardware with the time to solution. Note that we use only the TDP of the CPU and the GPUs for the energy to solution computations and we neglect the power consumption of other components which is typically considerably lower \citep[see e.g.][for a detailed energy-efficiency study of a computing cluster]{Cumming2014}. Energy savings of up to 98\% are possible compared to the current \texttt{LENSTOOL} software. The HPC techniques are thus friendly to the environment and lower the electricity bill of the computing cluster. While the use of the mixed precision algorithm further reduces the energy consumption in the case of the CPU and the GTX, its use increases the required energy for the P100 and the V100. This is due to the hybrid CPU/GPU approach in the mixed precision implementation for the GPUs. In the case of the GTX, the decrease in run time can handily offset the additional power consumption of the CPU, while this is not the case for the high-end GPUs.\\
\\

\begin{center}
	\begin{table*}
		\centering
		\begin{tabular}{c|c|c|c|c|c}
			& Hardware &Energy to solution & Energy to solution & Energy saved & Energy saved compared \\
			& TDP & Double Precision & Mixed Precision & Double $\to$ Mixed Precision & to \texttt{LENSTOOL} \\
			& (Watt) & (Joule) & (Joule) & (\%) & (\%) \\
			\hline
			\texttt{LENSTOOL} & 120 & 8016 & - & - & - \\
			CPU & 120 & ~972 &  888 & ~~9 & 88 / 89 \\
			GTX & 250 & 1700 & 585 & ~66 & 79 / 93 \\
			P100 & 300 & ~213 & 244 & -15 & 97 / 97 \\
			V100 & 300 & ~138 & 164 & -19 & 98 / 98 \\
		\end{tabular}
		\caption{ Energy comparison of the different gradient computation implementations for one run of the benchmark. We estimate the energy to solution by multiplying the Thermal Design Power (TDP) of the different hardware devices with the respective benchmark run times. In the case of the mixed precision GPU implementations, which use a hybrid CPU/GPU approach, we add the TDPs of the GPU and the CPU. The TDP values are taken from \citet{IntelCorporation2014} and \citet{NvidiaCorporation2016,NvidiaCorporation2017,NvidiaCorporation2017a}.  The last column shows the percentage of energy saved by using the double or mixed precision algorithm instead of the current \texttt{LENSTOOL} software. } \label{table_TDP}
	\end{table*}
\end{center}

It is possible to generalize this approach to other commonly used parametric lens models like the Navarro-Frenk-White (NFW) profile \citep{Navarro1997,Navarro1996} or the dual Pseudo Isothermal Elliptical mass distribution (dPIE) \citep[e.g.,][]{Eliasdottir2007,Kassiola1993}. However, this will lead to much more complicated and longer gradient computation algorithms than the one for the SIE studied in this paper. As a result, it is possible that the resulting error due to finite machine precision will become bigger, too. While the exact fraction of the image for which SP is accurate enough must be explicitly calculated and studied for the respective model, it is possible that it will be considerably lower than for the SIE. Therefore the performance gain from using mixed precision might shrink accordingly. A separate study will be necessary to determine whether the expected performance gain is worth the effort.\\
\\
Finally, it is necessary to discuss legal aspects of the GPU drivers provided by the graphics chip manufacturer, Nvidia. The drivers are required to use the hardware and their use is subject to certain terms and conditions, which Nvidia has recently updated. These new terms might be interpreted to prohibit the use of consumer-grade graphics cards like the GTX in computing clusters, which is the typical deployment in the scientific community. Thus the customers would be effectively forced to buy the much more expensive high-end GPUs, even though the cheaper graphics cards might be fully sufficient for the intended application. The authors of this paper are strongly concerned about this development, in particular given the limited financial budgets of academic research worldwide. Therefore they have contacted Nvidia and were informed in writing that Nvidia has no intentions to prohibit the use of the cheaper consumer-grade cards for the non-commercial purposes of researchers. The authors urge Nvidia to formalize this permission for scientific use by including it in the terms and conditions or preferably to remove this restriction altogether, thus allowing everyone to use the GPU which best suits their respective needs.

\section{Conclusion}\label{section_conclusion}
In this paper, we demonstrate the value of High Performance Computing techniques for strong lensing software. We study a performance-critical part of the widely used \texttt{LENSTOOL} lens modeling software, namely the deflection potential gradient computations of the $\chi^2$ calculation algorithm. We present and discuss an optimized CPU version with Advanced Vector Extensions and OpenMP and a GPU implementation in CUDA for a SIE lens model.\\
\\ 
In addition, we calculate the impact of finite machine precision on the strong lensing algorithm. We demonstrate for the SIE model that single precision is accurate enough for a large part of the image. We develop a mixed precision algorithm which allows us to use single precision for performance while computing critical parts of the image in double precision.\\
\\
Finally, we measure the computing performance for a galaxy cluster lens similar to the \textit{Hubble Frontier Fields}. We find that our HPC techniques accelerate the computation by an order of magnitude on CPUs and by up to two orders of magnitude on GPUs. In addition, they reduce the energy consumption by up to 98\%. The mixed precision approach delivers the best performance for every type of hardware while providing accurate results. It also permits to harness the full potential of a consumer-grade GPU, which can achieve a competitive benchmark performance for a small fraction of the monetary cost of a high-end GPU.

\section{Acknowledgments}
MR thanks Yves Revaz for fruitful discussions of GPU-accelerated computing. GF gratefully acknowledges support from the EPFL Facult\'{e} des Sciences de Base. This work was supported by EPFL through the use of the facilities of its Scientific IT and Application Support Center. The authors gratefully acknowledge the use of facilities of the Swiss National Supercomputing Centre (CSCS) and they thank Colin McMurtrie and Hussein Harake for their continued support. This research made use of matplotlib \citep{Hunter2007}, Inkscape, Astropy \citep{AstropyCollaboration2013}, TeX Live, Wolfram Alpha, and NASA's Astrophysics Data System.

\section*{References}
\bibliography{Astronomy_papers}

\appendix
\onecolumn
\section{Finite machine precision error in $\nabla \Psi_{\epsilon}$ computation for SIE}\label{appendix_a}
We compute the error for $\nabla \Psi_{\epsilon}$ for a SIE parametric model due to finite machine precision. To do so, we investigate each line of the $\nabla \Psi_{\epsilon}$ algorithm~\ref{algorithm:gradient_computation}, we compute the respective error due to finite machine precision, and we propagate the resulting errors into the next line of the algorithm. We use $\epsilon$ to denote the machine epsilon as defined in subsection~\ref{subsection_single_double_precision} and we have $\epsilon \approx 10^{-7}$ and $\epsilon \approx 10^{-16}$ for single and double precision, respectively. Thus we can neglect terms of the order $\mathcal{O}(\epsilon^{2})$ and higher. We make the assumption that the computer stores the result of one line of the algorithm, which typically corresponds to one line of code, in regular registers or memory. In addition, we assume that intermediate results, which occur while processing one line of the algorithm, are stored in extended precision registers, which are e.g. typically present in x87 Floating-Point Units (FPUs). As a result, we can neglect error contributions due to machine precision for these intermediate results. Note that this is no longer the case if we use Streaming SIMD Extension (SSE) or AVX registers, as these do not use extended precision. \\
\\
The Appendix is organized as follows: In \ref{appendix_subsection_error_rules}, we present the error propagation rules, in \ref{appendix_subsection_error_computation} we derive a mathematical expression for the error of $\nabla \Psi_{\epsilon}$ and we write it in a compact form by defining appropriate error variables, and in \ref{appendix_subsection_error_bounds} we compute upper bounds for the error.

\subsection{Error propagation rules}\label{appendix_subsection_error_rules}
We use the following error propagation rules which give upper limits on the propagated error:\\

\textbf{\\Addition:}
\begin{align}
	x \pm \epsilon a + y \pm \epsilon b = x + y \pm \epsilon (|a| + |b|)
\end{align}
\\
\textbf{Subtraction:}
\begin{align} 
	x \pm \epsilon a - y \pm \epsilon b = x - y \pm \epsilon (|a| + |b|)
\end{align}
\\
\textbf{Multiplication:}
\begin{align}
	(x \pm \epsilon a) (y \pm \epsilon b) &= x y \pm \epsilon |a y| \pm \epsilon |x b| \pm \mathcal{O}(\epsilon^{2})  \nonumber \\
	&= x y \pm \epsilon (|a y| + |x b|)
\end{align}
\\
\textbf{Division:}
\begin{align}
	\frac{x \pm \epsilon a}{y \pm \epsilon b} = \frac{x}{y} \pm \epsilon \frac{|a y| + |b x|}{y^{2} \pm \epsilon b y}
\end{align}
\\
Proof:
\begin{align}
	\frac{x \pm \epsilon a}{y \pm \epsilon b} - \frac{x}{y} &= \frac{(x \pm \epsilon a) y - x (y \pm \epsilon b)}{y^{2} \pm \epsilon b y} \nonumber\\
	&= \frac{x y \pm \epsilon a y - x y \pm \epsilon b x}{y^{2} \pm \epsilon b y} \nonumber \\
	&= \pm \epsilon \frac{|a y| + |b x|}{y^{2} \pm \epsilon b y } \nonumber
\end{align}
\\
\textbf{General, infinitely differentiable function $f(x)$:}\\
\\
We can use the Taylor expansion to first order,
\begin{align}
	f(x \pm \epsilon a) = f(x) \pm \epsilon a f^{\prime}(x),
\end{align}
if the contribution from higher order terms is negligible:
\begin{align}
	\frac{f^{n}(x) a^{n} \epsilon^{n} }{f^{\prime}(x) a \epsilon n!} \approx 0~~ \forall n > 1, \nonumber
\end{align}
where $f^{n}(x)$ denotes the $n$-th derivative.

\subsection{Error computation}\label{appendix_subsection_error_computation}
We will denote a result $x$ stored in a regular register or memory with $\texttt{stored}(x)$. We want to derive an upper limit on the final error, so we will assume that each of these storage operations produces an error, $\texttt{stored}(x) = x \pm \epsilon x$, and we propagate these errors. Note that in practice the storing of results does not necessarily produce an error and, since the storing error is basically due to a rounding operation, errors from different storing operations can cancel each other.\\
\\
We compute now the machine precision error for one SIE lens at pixel $(\theta_{1},\theta_{2})$:

\begin{flalign}
\Delta \theta_{1} &= \theta_{1} \pm \epsilon \theta_{1} - (\theta_{\text{center},1} \pm \epsilon \theta_{\text{center},1}) \nonumber \\
&= \theta_{1} - \theta_{\text{center},1} \pm \epsilon (|\theta_{1}| + |\theta_{\text{center},1}|). &
\end{flalign}

\begin{flalign}
\texttt{stored}(\Delta \theta_{1}) &=	\theta_{1} - \theta_{\text{center},1} \pm \epsilon (|\theta_{1}| + |\theta_{\text{center},1}|) \pm \epsilon (|\theta_{1} - \theta_{\text{center},1}|) \pm \mathcal{O}(\epsilon^{2}) \nonumber \\ 
&= \theta_{1} - \theta_{\text{center},1} \pm \epsilon (|\theta_{1}| + |\theta_{\text{center},1}| + |\theta_{1} - \theta_{\text{center},1}|) \nonumber \\
&= \Delta \theta_{1,t} \pm \epsilon A_{1}.&
\end{flalign}
In the last line, we introduced the true, error-free value of $\Delta \theta_{1}$, $\Delta \theta_{1,t} = \theta_{1} - \theta_{\text{center},1}$. In addition, we implicitly defined the error variable $A_{1}$, which contains all the terms which contribute to the error.\\
\\
\begin{flalign}
\Delta \theta_{2} &= \theta_{2} \pm \epsilon \theta_{2} - (\theta_{\text{center},2} \pm \epsilon \theta_{\text{center},2}) \nonumber \\
&= \theta_{2} - \theta_{\text{center},2} \pm \epsilon (|\theta_{2}| + |\theta_{\text{center},2}|). &
\end{flalign}

\begin{flalign}
\texttt{stored}(\Delta \theta_{2}) &=	\theta_{2} - \theta_{\text{center},2} \pm \epsilon (|\theta_{2}| + |\theta_{\text{center},2}|) \pm \epsilon (|\theta_{2} - \theta_{\text{center},2}|) \pm \mathcal{O}(\epsilon^{2}) \nonumber \\ 
&= \theta_{2} - \theta_{\text{center},2} \pm \epsilon (|\theta_{2}| + |\theta_{\text{center},2}| + |\theta_{2} - \theta_{\text{center},2}|) \nonumber \\
&= \Delta \theta_{2,t} \pm \epsilon A_{2},&
\end{flalign}
where we again implicity defined $\Delta \theta_{2,t}$ and $A_{2}$.\\
\\
\begin{flalign}
	\Delta \theta_{1}^{\prime} &= (\Delta \theta_{1,t} \pm \epsilon A_{1}) \cos(\Phi \pm \epsilon \Phi) + (\Delta \theta_{2,t} \pm \epsilon A_{2}) \sin(\Phi \pm \epsilon \Phi) \nonumber \\
	&= (\Delta \theta_{1,t} \pm \epsilon A_{1}) [\cos(\Phi) \pm \sin(\Phi) \epsilon \Phi] + (\Delta \theta_{2,t} \pm \epsilon A_{2}) [\sin(\Phi) \pm \cos(\Phi) \epsilon \Phi]. &
\end{flalign}
As we have $\sin(\Phi) \leq 1$ and $\cos(\Phi) \leq 1$, we can obtain an upper bound on the error by replacing the respective sine and cosine expressions in the parts which contribute to the error with $1$: 
\begin{flalign}
	\Delta \theta_{1}^{\prime} &= (\Delta \theta_{1,t} \pm \epsilon A_{1}) [\cos(\Phi) \pm \epsilon \Phi] + (\Delta \theta_{2,t} \pm \epsilon A_{2}) [\sin(\Phi) \pm \epsilon \Phi] \nonumber \\
	&= \Delta \theta_{1,t} \cos(\Phi) \pm \epsilon(|A_{1} \cos(\Phi)| + |\Delta\theta_{1,t} \Phi|) + \Delta \theta_{2,t} \sin(\Phi) \pm \epsilon (|A_{2} \sin(\Phi)| + |\Delta \theta_{2,t} \Phi|) + \mathcal{O}(\epsilon^{2}) \nonumber \\
	&= \Delta \theta_{1,t} \cos(\Phi) \pm \epsilon(|A_{1}| + |\Delta\theta_{1,t} \Phi|) + \Delta \theta_{2,t} \sin(\Phi)  \pm \epsilon (|A_{2}| + |\Delta \theta_{2,t} \Phi|) \nonumber \\
	&= \Delta \theta_{1,r} \pm \epsilon (|A_{1}| + |A_{2}| + |\Delta\theta_{1,t} \Phi| + |\Delta \theta_{2,t} \Phi|). &
\end{flalign}
In the last line we implicitly defined the true value after rotation, $\Delta\theta_{1,r}$.\\
\\
\begin{flalign}
	\texttt{stored}(\Delta\theta_{1}^{\prime}) &= \Delta \theta_{1,r} \pm \epsilon (|A_{1}| + |A_{2}| + |\Delta \theta_{1,t} \Phi| + |\Delta \theta_{2,t} \Phi| + |\Delta \theta_{1,t} \cos(\Phi)| + |\Delta \theta_{2,t} \sin(\Phi)|) + \mathcal{O}(\epsilon^{2}) \nonumber \\
	&= \Delta \theta_{1,r} \pm \epsilon (|A_{1}| + |A_{2}| + |\Delta \theta_{1,t} \Phi| + |\Delta \theta_{2,t} \Phi| + |\Delta \theta_{1,t}| + |\Delta \theta_{2,t}|) \nonumber \\
	&= \Delta \theta_{1,r} \pm \epsilon B, &
\end{flalign}
where we again replaced sine and cosine with 1 and implicitly defined the error variable $B$.\\
\\
\\
Similarly, we obtain for $\Delta \theta_{2}^{\prime}:$
\begin{flalign}
\Delta \theta_{2}^{\prime} &= (\Delta \theta_{2,t} \pm \epsilon A_{2}) [\cos(\Phi) \pm \epsilon \Phi] - (\Delta \theta_{1,t} \pm \epsilon A_{1}) [\sin(\Phi) \pm \epsilon \Phi] \nonumber \\
&= \Delta \theta_{2,t} \cos(\Phi) \pm \epsilon(|A_{2}| + |\Delta\theta_{2,t} \Phi|) - \Delta \theta_{1,t} \sin(\Phi)  \pm \epsilon (|A_{1}| + |\Delta \theta_{1,t} \Phi|) + \mathcal{O}(\epsilon^{2}) \nonumber \\
&= \Delta \theta_{2,r} \pm \epsilon (|A_{1}| + |A_{2}| + |\Delta\theta_{1,t} \Phi| + |\Delta \theta_{2,t} \Phi|). &
\end{flalign}
In the last line we implicitly defined the true value after rotation, $\Delta\theta_{2,r}$.\\
\\
\begin{flalign}
	\texttt{stored}(\Delta \theta_{2}^{\prime}) &= \Delta \theta_{2,r} \pm \epsilon B.&
\end{flalign}\\
\\
In this appendix, we denote the pseudo-ellipticity of the deflection potential with $p$ instead of $\epsilon$ to avoid confusion with the machine epsilon. We further define $p_{\star} = 1-p$, $p_{\dagger} = 1 + p$ and we obtain:
\begin{flalign}
	R &= \sqrt{(\Delta \theta_{1,r} \pm \epsilon B)^{2} (1 - p \pm \epsilon p_{\star} ) + (\Delta \theta_{2,r} \pm \epsilon B)^{2} (1+p \pm \epsilon p_{\dagger})  }  \nonumber \\
	&= \sqrt{(\Delta \theta_{1,r}^{2} \pm \epsilon |2 \Delta \theta_{1,r} B|) (1 - p \pm \epsilon p_{\star} ) + (\Delta \theta_{2,r}^{2} \pm \epsilon |2 \Delta \theta_{2,r} B|) (1+p \pm \epsilon p_{\dagger}) + \mathcal{O}(\epsilon^{2}) }  \nonumber \\
	&= \sqrt{ \Delta \theta_{1,r}^{2} (1-p) \pm \epsilon (|\Delta \theta_{1,r}^{2} p_{\star}| + |2 \Delta \theta_{1,r} B (1-p)|)  + \Delta \theta_{2,r}^{2} (1+p) \pm \epsilon  (|\Delta \theta_{2,r}^{2} p_{\dagger}| + |2 \Delta \theta_{2,r} B (1+p)|) + \mathcal{O}(\epsilon^{2})} \nonumber \\
	&= \sqrt{ \Delta \theta_{1,r}^{2} (1-p) + \Delta \theta_{2,r}^{2} (1+p) \pm \epsilon (|\Delta \theta_{1,r}^{2} p_{\star}| +  |\Delta \theta_{2,r}^{2} p_{\dagger}| + |2 \Delta \theta_{1,r} B (1-p)| + |2 \Delta \theta_{2,r} B (1+p)| ) }. &
\end{flalign}
We define the true value of $R$,
\begin{flalign}
	R_{t} &= \sqrt{\Delta \theta_{1,r}^{2} (1-p) + \Delta \theta_{2,r}^{2} (1+p)}, &
\end{flalign}
and use a Taylor expansion to obtain
\begin{flalign}
	R &= R_{t} \pm \epsilon \frac{|\Delta \theta_{1,r}^{2} p_{\star}| + |\Delta \theta_{2,r}^{2} p_{\dagger}| + |2 \Delta \theta_{1,r} B (1-p)| + |2 \Delta \theta_{2,r} B (1+p)|}{2 |R_{t}|}. &
\end{flalign}\\
\\
\begin{flalign}
	\texttt{stored}(R) &= R_{t} \pm \epsilon \bigg(\frac{|\Delta \theta_{1,r}^{2} p_{\star}| + |\Delta \theta_{2,r}^{2} p_{\dagger}| + |2 \Delta \theta_{1,r} B (1-p)| + |2 \Delta \theta_{2,r} B (1+p)|}{2 |R_{t}|} + |R_{t}| \bigg) + \mathcal{O}(\epsilon^{2}) \nonumber \\
	&= R_{t} \pm \epsilon C, &
\end{flalign}
where we implicitly defined the error variable $C$.\\
\\
\begin{flalign}
	\nabla \Psi_{\epsilon,1} &= (1 - p \pm \epsilon p_{\star}) (b_{0} \pm \epsilon b_{0}) \frac{\Delta \theta_{1,r} \pm \epsilon B}{R_{t} \pm \epsilon C} \nonumber \\
	&= [(1-p) b_{0} \pm \epsilon (2 b_{0} p_{\star})] \frac{\Delta \theta_{1,r} \pm \epsilon B}{R_{t} \pm \epsilon C} + \mathcal{O}(\epsilon^{2}) \nonumber \\
	&= [(1-p) b_{0} \pm \epsilon (2 b_{0} p_{\star})] \bigg[\frac{\Delta \theta_{1,r}}{R_{t}} \pm \epsilon \frac{|B R_{t}| + |C \Delta \theta_{1,r}|}{R_{t}^{2} \pm \epsilon |C R_{t}|}\bigg] \nonumber \\
	&= (1-p) b_{0} \frac{\Delta \theta_{1,r}}{R_{t}} \pm \epsilon \bigg[\Big|2 b_{0} p_{\star} \frac{\Delta \theta_{1,r}}{R_{t}}\Big| + \Big|(1-p) b_{0} \frac{|B R_{t}| + |C \Delta \theta_{1,r}|}{R_{t}^{2} \pm \epsilon |C R_{t}|}\Big|\bigg] + \mathcal{O}(\epsilon^{2}) \nonumber \\
	&= \nabla \Psi_{\epsilon,t,1} \pm \epsilon \bigg[|2 \nabla \Psi_{\epsilon,t,1}| + \Big|(1-p) b_{0} \frac{|B R_{t}| + |C \Delta \theta_{1,r}|}{R_{t}^{2} \pm \epsilon |C R_{t}|}\Big|\bigg], &
\end{flalign}
where we implicitly defined the true value of the gradient, $\nabla \Psi_{\epsilon,t,1}$.\\
\\
\begin{flalign}
	\texttt{stored}(\nabla \Psi_{\epsilon,1}) &= \nabla \Psi_{\epsilon,t,1} \pm \epsilon \bigg[|3 \nabla \Psi_{\epsilon,t,1}| + \Big|(1-p) b_{0} \frac{|B R_{t}| + |C \Delta \theta_{1,r}|}{R_{t}^{2} \pm \epsilon |C R_{t}|}\Big|\bigg] + \mathcal{O}(\epsilon^{2}) \nonumber \\
	&= \nabla \Psi_{\epsilon,t,1} \pm \epsilon D_{1},&
\end{flalign}
where we implicitly defined the error variable $D_{1}$.\\
\\
\\
\begin{flalign}
	\nabla \Psi_{\epsilon,2} &= (1 + p \pm \epsilon p_{\dagger}) (b_{0} \pm \epsilon b_{0}) \frac{\Delta \theta_{2,r} \pm \epsilon B}{R_{t} \pm \epsilon C} \nonumber \\
&= [(1+p) b_{0} \pm \epsilon (2 b_{0} p_{\dagger})] \frac{\Delta \theta_{2,r} \pm \epsilon B}{R_{t} \pm \epsilon C} + \mathcal{O}(\epsilon^{2}) \nonumber \\
&= [(1+p) b_{0} \pm \epsilon (2 b_{0} p_{\dagger})] \bigg[\frac{\Delta \theta_{2,r}}{R_{t}} \pm \epsilon \frac{|B R_{t}| + |C \Delta \theta_{2,r}|}{R_{t}^{2} \pm \epsilon |C R_{t}|}\bigg] \nonumber \\
&= (1+p) b_{0} \frac{\Delta \theta_{2,r}}{R_{t}} \pm \epsilon \bigg[\Big|2 b_{0} p_{\dagger} \frac{\Delta \theta_{2,r}}{R_{t}}\Big| + \Big|(1+p) b_{0} \frac{|B R_{t}| + |C \Delta \theta_{2,r}|}{R_{t}^{2} \pm \epsilon |C R_{t}|}\Big|\bigg] + \mathcal{O}(\epsilon^{2}) \nonumber \\
&= \nabla \Psi_{\epsilon,t,2} \pm \epsilon \bigg[|2 \nabla \Psi_{\epsilon,t,2}| + \Big|(1+p) b_{0} \frac{|B R_{t}| + |C \Delta \theta_{2,r}|}{R_{t}^{2} \pm \epsilon |C R_{t}|}\Big|\bigg], &
\end{flalign}
where we implicitly defined the true value of the gradient, $\nabla \Psi_{\epsilon,t,2}$.\\
\\
\begin{flalign}
\texttt{stored}(\nabla \Psi_{\epsilon,2}) &= \nabla \Psi_{\epsilon,t,2} \pm \epsilon \bigg[|3 \nabla \Psi_{\epsilon,t,2}| + \Big|(1+p) b_{0} \frac{|B R_{t}| + |C \Delta \theta_{2,r}|}{R_{t}^{2} \pm \epsilon |C R_{t}|}\Big|\bigg] + \mathcal{O}(\epsilon^{2}) \nonumber \\
&= \nabla \Psi_{\epsilon,t,2} \pm \epsilon D_{2},&
\end{flalign}
where we implicitly defined the error variable $D_{2}$.\\
\\
\begin{flalign}
	\nabla \Psi_{\epsilon,1}^{\prime} &= (\nabla \Psi_{\epsilon,t,1} \pm \epsilon D_{1}) \cos(-\Phi \pm \epsilon \Phi) + (\nabla \Psi_{\epsilon,t,2} \pm \epsilon D_{2}) \sin(-\Phi \pm \epsilon \Phi) \nonumber \\
	&= (\nabla \Psi_{\epsilon,t,1} \pm \epsilon D_{1}) [\cos(-\Phi) \pm \sin(-\Phi) \epsilon \Phi] + (\nabla \Psi_{\epsilon,t,2} \pm \epsilon D_{2}) [\sin(-\Phi) \pm \cos(-\Phi) \epsilon \Phi]. &
\end{flalign}
As we have $\sin(\Phi) \leq 1$ and $\cos(\Phi) \leq 1$, we can obtain an upper bound on the error by replacing the respective sine and cosine expressions in the parts which contribute to the error with $1$: 
\begin{flalign}
	\nabla \Psi_{\epsilon,1}^{\prime} &= (\nabla \Psi_{\epsilon,t,1} \pm \epsilon D_{1}) [\cos(-\Phi) \pm \epsilon \Phi] + (\nabla \Psi_{\epsilon,t,2} \pm \epsilon D_{2}) [\sin(-\Phi) \pm \epsilon \Phi] \nonumber \\
	&= \nabla \Psi_{\epsilon,t,1} \cos(-\Phi) \pm \epsilon (|\nabla \Psi_{\epsilon,t,1} \Phi| + |D_{1} \cos(-\Phi)|) + \nabla \Psi_{\epsilon,t,2} \sin(-\Phi) \pm \epsilon (|\nabla \Psi_{\epsilon,t,2} \Phi| + |D_{2} \sin(-\Phi)|) + \mathcal{O}(\epsilon^{2}) \nonumber \\
	&= \nabla \Psi_{\epsilon,t,1} \cos(-\Phi) \pm \epsilon (|\nabla \Psi_{\epsilon,t,1} \Phi| + |D_{1}|) + \nabla \Psi_{\epsilon,t,2} \sin(-\Phi) \pm \epsilon (|\nabla \Psi_{\epsilon,t,2} \Phi| + |D_{2}|) \nonumber \\
	&=  \nabla \Psi_{\epsilon,r,1} \pm \epsilon (|\nabla \Psi_{\epsilon,t,1} \Phi| + |\nabla \Psi_{\epsilon,t,2} \Phi| + |D_{1}| + |D_{2}|), &
\end{flalign}
where we implicitly defined the true value of the first gradient component after rotation, $\nabla \Psi_{\epsilon,r,1}$.\\
\\
\\
We use the relation
\begin{flalign}
	|\nabla \Psi_{\epsilon,t,1} \cos(-\Phi) + \nabla \Psi_{\epsilon,t,2} \sin(-\Phi)| \leq |\nabla \Psi_{\epsilon,t,1} \cos(-\Phi)| + |\nabla \Psi_{\epsilon,t,2} \sin(-\Phi)| \leq |\nabla \Psi_{\epsilon,t,1}| + |\nabla \Psi_{\epsilon,t,2}| &&
\end{flalign}
to obtain:
\begin{flalign}
	\texttt{stored}(\nabla \Psi_{\epsilon,1}^{\prime}) &= \nabla \Psi_{\epsilon,r,1} \pm \epsilon (|\nabla \Psi_{\epsilon,t,1} \Phi| + |\nabla \Psi_{\epsilon,t,2} \Phi| + |D_{1}| + |D_{2}| + |\nabla \Psi_{\epsilon,t,1}| + |\nabla \Psi_{\epsilon,t,2}|) + \mathcal{O}(\epsilon^{2}) \nonumber \\
	&= \nabla \Psi_{\epsilon,r,1} \pm \epsilon F, &
\end{flalign}
where we implicitly defined the error variable $F$.\\
\\
\begin{flalign}
\nabla \Psi_{\epsilon,2}^{\prime} &= (\nabla \Psi_{\epsilon,t,2} \pm \epsilon D_{2}) \cos(-\Phi \pm \epsilon \Phi) - (\nabla \Psi_{\epsilon,t,1} \pm \epsilon D_{1}) \sin(-\Phi \pm \epsilon \Phi) \nonumber \\
&= (\nabla \Psi_{\epsilon,t,2} \pm \epsilon D_{2}) [\cos(-\Phi) \pm \sin(-\Phi) \epsilon \Phi] - (\nabla \Psi_{\epsilon,t,1} \pm \epsilon D_{1}) [\sin(-\Phi) \pm \cos(-\Phi) \epsilon \Phi]. &
\end{flalign}
As we have $\sin(\Phi) \leq 1$ and $\cos(\Phi) \leq 1$, we can obtain an upper bound on the error by replacing the respective sine and cosine expressions in the parts which contribute to the error with $1$: 
\begin{flalign}
\nabla \Psi_{\epsilon,2}^{\prime} &= (\nabla \Psi_{\epsilon,t,2} \pm \epsilon D_{2}) [\cos(-\Phi) \pm \epsilon \Phi] - (\nabla \Psi_{\epsilon,t,1} \pm \epsilon D_{1}) [\sin(-\Phi) \pm \epsilon \Phi] \nonumber \\
&= \nabla \Psi_{\epsilon,t,2} \cos(-\Phi) \pm \epsilon (|\nabla \Psi_{\epsilon,t,2} \Phi| + |D_{2} \cos(-\Phi)|) - \nabla \Psi_{\epsilon,t,1} \sin(-\Phi) \pm \epsilon (|\nabla \Psi_{\epsilon,t,1} \Phi| + |D_{1} \sin(-\Phi)|) + \mathcal{O}(\epsilon^{2}) \nonumber \\
&= \nabla \Psi_{\epsilon,t,2} \cos(-\Phi) \pm \epsilon (|\nabla \Psi_{\epsilon,t,2} \Phi| + |D_{2}|) - \nabla \Psi_{\epsilon,t,1} \sin(-\Phi) \pm \epsilon (|\nabla \Psi_{\epsilon,t,1} \Phi| + |D_{1}|) \nonumber \\
&=  \nabla \Psi_{\epsilon,r,2} \pm \epsilon (|\nabla \Psi_{\epsilon,t,1} \Phi| + |\nabla \Psi_{\epsilon,t,2} \Phi| + |D_{1}| + |D_{2}|), &
\end{flalign}
where we implicitly defined the true value of the second gradient component after rotation, $\nabla \Psi_{\epsilon,r,2}$.\\
\\
\\
We use the relation
\begin{flalign}
|\nabla \Psi_{\epsilon,t,2} \cos(-\Phi) - \nabla \Psi_{\epsilon,t,1} \sin(-\Phi)| \leq |\nabla \Psi_{\epsilon,t,2} \cos(-\Phi)| + |\nabla \Psi_{\epsilon,t,1} \sin(-\Phi)| \leq |\nabla \Psi_{\epsilon,t,2}| + |\nabla \Psi_{\epsilon,t,1}| &&
\end{flalign}
to obtain:
\begin{flalign}
\texttt{stored}(\nabla \Psi_{\epsilon,2}^{\prime}) &= \nabla \Psi_{\epsilon,r,2} \pm \epsilon (|\nabla \Psi_{\epsilon,t,1} \Phi| + |\nabla \Psi_{\epsilon,t,2} \Phi| + |D_{1}| + |D_{2}| + |\nabla \Psi_{\epsilon,t,1}| + |\nabla \Psi_{\epsilon,t,2}|) + \mathcal{O}(\epsilon^{2}) \nonumber \\
&= \nabla \Psi_{\epsilon,r,2} \pm \epsilon F. &
\end{flalign}
\\
\\
As a result, the total error of one computed $\nabla \Psi_{\epsilon}$ for one pixel $(\theta_{1},\theta_{2})$ due to finite machine precision is $\epsilon F$ for both gradient components.
\\
\\

\subsection{Upper error bounds for cluster- and galaxy-scale SIE lenses}\label{appendix_subsection_error_bounds}
\subsubsection{Centered SIE lenses}
Let us consider a single lens at the origin of a two dimensional image plane coordinate system,
\begin{equation}
	(\theta_{\text{center},1},\theta_{\text{center},2}) = (0,0). 
\end{equation} 
We assume that the point $(\theta_{1},\theta_{2})$ for which we compute $\nabla \Psi_{\epsilon}$ lies on the $\theta_{1}$-axis. We assume that we can do so without loss of generality, as this can be achieved by a simple rotation of the coordinate system. This simplifies the expression for the $A$ terms to
\begin{align}
	A_{1} &= 2|\theta_{1}|, \nonumber \\
	A_{2} &= 0.
\end{align}
We want to maximize the errors to obtain an upper bound. Therefore we maximize the angle $\Phi$, which always appears as an error increasing factor in the error variables. Due to the symmetry of an ellipse, the largest value is $\Phi = \pi$. As a result, we have
\begin{align}
	B = 2|\theta_{1}| + \pi |\theta_{1}| + |\theta_{1}| = (3+\pi) |\theta_{1}|.
\end{align}
Note that our coordinate system is now rotated by 180 degrees, so we have 
\begin{align}
	\Delta \theta_{1} \to - \Delta \theta_{1}^{\prime}, \nonumber \\
	\Delta \theta_{2} \to - \Delta \theta_{2}^{\prime}.
\end{align}
Next, we note that the pseudo-ellipticity $p$ is typically small and that it appears in the error variables in connection with $\Delta \theta_{1}^{\prime}$ as a factor $1-p$ and in connection with $\Delta \theta_{2}^{\prime}$ as a factor $1+p$. Therefore we will minimize it and assume $p = 0$. Thus we have
\begin{align}
	C = \frac{|\theta_{1}^{2}| + (6+2\pi)|\theta_{1}^{2}|}{2|\theta_{1}|} + |\theta_{1}| = (4.5 + \pi) |\theta_{1}|.
\end{align}
The lensing effect will be maximal for a source at high redshift, so we assume $D_{\text{OS}}/D_{\text{LS}} = 1$ and thus we have
\begin{align}
	|\nabla \Psi_{\epsilon,1}| &= |(\nabla \psi)_{1}| = \theta_{\text{E}}, \nonumber \\
	|\nabla \Psi_{\epsilon,2}| &= |(\nabla \psi)_{2}| = 0, \\
	b_{0} &= \theta_{\text{E}},
\end{align}
and thus
\begin{align}
	D_{1} &\approx 3 \theta_{\text{E}} + \theta_{\text{E}} \frac{(3+\pi) |\theta_{1}|^{2} + (4.5 + \pi) |\theta_{1}|^{2}}{|\theta_{1}|^{2}} = (10.5 + 2\pi) \theta_{\text{E}}, \nonumber \\
	D_{2} &\approx \theta_{\text{E}} \frac{(3+\pi) |\theta_{1}|^{2}}{|\theta_{1}|^{2}} = (3+\pi) \theta_{\text{E}}.
\end{align}
We now rotate the coordinate system by -180 degrees,
\begin{align}
	\nabla \Psi_{\epsilon,1} \to - \nabla \Psi_{\epsilon,1}^{\prime}, \nonumber \\
	\nabla \Psi_{\epsilon,2} \to - \nabla \Psi_{\epsilon,2}^{\prime}, 
\end{align}
and we obtain 
\begin{align}
	F = \pi \theta_{E} + (10.5 + 2\pi) \theta_{\text{E}} + (3+\pi) \theta_{\text{E}} + \theta_{\text{E}} \approx 27 \theta_{\text{E}}.
\end{align}
As a result, we have for a cluster-scale halo with $\theta_{\text{E}} = 20~\text{arcsec}$
\begin{equation}
	F_{\text{cluster-scale}} = 540~\text{arcsec}
\end{equation}
and for a galaxy-scale halo with $\theta_{\text{E}} = 0.2~\text{arcsec}$
\begin{equation}
	F_{\text{galaxy-scale}} = 5.4~\text{arcsec}.
\end{equation}
For single and double precision, we have respectively $\epsilon_{\text{SP}} \approx 1.2 \times 10^{-7}$ and $\epsilon_{\text{DP}} \approx 2.2 \times 10^{-16}$, and thus the upper error bounds
\begin{align}
	\epsilon_{\text{SP}} F_{\text{cluster-scale}} &\approx 6.5 \times 10^{-5}~\text{arcsec}, \nonumber \\
	\epsilon_{\text{SP}} F_{\text{galaxy-scale}} &\approx 6.5 \times 10^{-7}~\text{arcsec}, \\
	\epsilon_{\text{DP}} F_{\text{cluster-scale}} &\approx 1.2 \times 10^{-13}~\text{arcsec}, \nonumber \\
	\epsilon_{\text{DP}} F_{\text{galaxy-scale}} &\approx 1.2 \times 10^{-15}~\text{arcsec}.
\end{align}
The computed gradients for each halo are finally added up to obtain the total gradient, 
\begin{equation}
	\nabla \Psi_{\epsilon,i} = \sum_{k} \nabla \Psi_{\epsilon,i,k}^{\prime},
\end{equation}
and as a result, the respective errors are combined as well. However, the respective errors can have different signs and magnitudes, so we expect to see some error cancellation. We estimate the total gradient error in the following way: We neglect the contribution from the galaxy-scale halos and we add the respective upper error bounds of the cluster-scale halos. Neglecting the galaxy-scale lenses is justified, because first, their absolute errors are two orders of magnitude smaller than those of the cluster-scale halos, and second, we add many of these halos which are typically scattered throughout the image, so we expect significant error cancellation effects. We are left with typically two cluster-scale halos. The error contribution from these halos will depend on their respective parameters. To obtain an upper bound, we will add up the respective upper bounds on the gradient, so we have
\begin{align}
	\Delta (\nabla \Psi_{\epsilon,i})_{SP} &\approx 1.3 \times 10^{-4}~\text{arcsec}, \nonumber \\
	\Delta (\nabla \Psi_{\epsilon,i})_{DP} &\approx 2.4 \times 10^{-13}~\text{arcsec}.
\end{align}

\subsubsection{General SIE lenses}
In the previous part, we implicitly assumed that the finite machine precision error is invariant under translations and rotations of the coordinate system. As a result, it was sufficient to compute the error for a single centered SIE lens and we could use the result to derive the total error for the cluster lens system. However, we now show that this assumed invariance only holds approximately and only far away from the lens center.\\
\\  
Let us consider a HST ACS image. We let the origin of the coordinate system coincide with the first pixel of the image in the lower left corner. Consequently all pixel values are positive, so we have
\begin{align}
	|\theta_{i}| \leq |\theta_{i} - \theta_{\text{center},i}| + |\theta_{\text{center},i}|
\end{align} 
and thus the upper bounds for the $A$ terms are
\begin{align}
A_{1} &= 2|\Delta \theta_{1}| + 2|\theta_{\text{center},1}|, \nonumber \\
A_{2} &= 2|\Delta \theta_{2}| + 2|\theta_{\text{center},2}|,
\end{align}
where 
\begin{align}
	|\Delta \theta_{i}| = |\theta_{i} - \theta_{\text{center},i}|.
\end{align} 
We want to maximize the errors to obtain an upper bound. Therefore we maximize the angle $\Phi$, which always appears as an error increasing factor in the error variables. Due to the symmetry of an ellipse, the largest value is $\Phi = \pi$. As a result, we have
\begin{align}
B &= 2|\Delta\theta_{1}| + 2|\theta_{\text{center},1}| + 2|\Delta\theta_{2}| + 2|\theta_{\text{center},2}| + \pi |\Delta\theta_{1}| + \pi |\Delta\theta_{2}| + |\Delta\theta_{1}| + |\Delta\theta_{2}| \nonumber \\
&= (3+\pi) |\Delta\theta_{1}| + (3+\pi) |\Delta\theta_{2}| + 2|\theta_{\text{center},1}| + 2|\theta_{\text{center},2}|.
\end{align}
Note that our coordinate system is now rotated by 180 degrees, so we have 
\begin{align}
\Delta \theta_{1} \to - \Delta \theta_{1}^{\prime}, \nonumber \\
\Delta \theta_{2} \to - \Delta \theta_{2}^{\prime}.
\end{align}
Next, we note that the pseudo-ellipticity $p$ is typically small and that it appears in the error variables in connection with $\Delta \theta_{1}^{\prime}$ as a factor $1-p$ and in connection with $\Delta \theta_{2}^{\prime}$ as a factor $1+p$. Therefore we will minimize it and assume $p = 0$. Thus we have
\begin{align}
C &= \frac{|\Delta\theta_{1}^{2}| + |\Delta\theta_{2}^{2}| + (6+2\pi)|\Delta\theta_{1}^{2}| + (6+2\pi)|\Delta\theta_{2}^{2}|  + (12+4\pi)|\Delta\theta_{1}||\Delta\theta_{2}| + 4(|\Delta\theta_{1}| + |\Delta\theta_{2}|)(|\theta_{\text{center},1}| + |\theta_{\text{center},2}|) }{2\sqrt{|\Delta\theta_{1}^{2}| + |\Delta\theta_{2}^{2}|}} \nonumber \\
&\hphantom{=\,\,} + \Big|\sqrt{|\Delta\theta_{1}^{2}| + |\Delta\theta_{2}^{2}|}  \,  \Big| \nonumber \\
&= \frac{(12+4\pi)|\Delta\theta_{1}||\Delta\theta_{2}| + 4(|\Delta\theta_{1}| + |\Delta\theta_{2}|)(|\theta_{\text{center},1}| + |\theta_{\text{center},2}|)}{2\sqrt{|\Delta\theta_{1}^{2}| + |\Delta\theta_{2}^{2}|}} + (4.5 + \pi) \Big|\sqrt{|\Delta\theta_{1}^{2}| + |\Delta\theta_{2}^{2}|}  \,  \Big|.
\end{align}
The lensing effect will be maximal for a source at high redshift, so we assume $D_{\text{OS}}/D_{\text{LS}} = 1$ and thus we have
\begin{align}
|\nabla \Psi_{\epsilon,1}| &= |(\nabla \psi)_{1}| \leq \theta_{\text{E}}, \nonumber \\
|\nabla \Psi_{\epsilon,2}| &= |(\nabla \psi)_{2}| \leq \theta_{\text{E}}, \\
b_{0} &= \theta_{\text{E}},
\end{align}
and thus
\begin{align}
D_{1} &\approx 3 \theta_{\text{E}} + \theta_{\text{E}} \frac{(3+\pi) (|\Delta\theta_{1}| + |\Delta\theta_{2}|) \sqrt{|\Delta\theta_{1}^{2}| + |\Delta\theta_{2}^{2}|} + 2(|\theta_{\text{center},1}| + |\theta_{\text{center},2}|) \sqrt{|\Delta\theta_{1}^{2}| + |\Delta\theta_{2}^{2}|}}{|\Delta\theta_{1}|^{2} + |\Delta\theta_{2}|^{2}}  \nonumber \\
&\hphantom{=\,\,} + \theta_{\text{E}} \frac{(4.5 + \pi) |\Delta\theta_{1}|  \sqrt{|\Delta\theta_{1}^{2}| + |\Delta\theta_{2}^{2}|}}{|\Delta\theta_{1}|^{2} + |\Delta\theta_{2}|^{2}} + \theta_{\text{E}} \frac{(6+2\pi)|\Delta\theta_{1}^{2}||\Delta\theta_{2}| + 2(|\Delta\theta_{1}^{2}| + |\Delta\theta_{1}||\Delta\theta_{2}|)(|\theta_{\text{center},1}| + |\theta_{\text{center},2}|)}{(|\Delta\theta_{1}^{2}| + |\Delta\theta_{2}^{2}|)^{\frac{3}{2}}}\nonumber \\
&= \theta_{\text{E}} \bigg[3 + \frac{(3 + \pi) (|\Delta\theta_{1}| + |\Delta\theta_{2}|) + 2(|\theta_{\text{center},1}| + |\theta_{\text{center},2}|) + (4.5 + \pi)|\Delta\theta_{1}|}{\sqrt{|\Delta\theta_{1}^{2}| + |\Delta\theta_{2}^{2}|}}   \nonumber \\
&\hphantom{=\,\,} +   \frac{(6+2\pi)|\Delta\theta_{1}^{2}||\Delta\theta_{2}| + 2(|\Delta\theta_{1}^{2}| + |\Delta\theta_{1}||\Delta\theta_{2}|)(|\theta_{\text{center},1}| + |\theta_{\text{center},2}|)}{(|\Delta\theta_{1}^{2}| + |\Delta\theta_{2}^{2}|)^{\frac{3}{2}}}  \bigg], 
\end{align}
and
\begin{align}
D_{2} &\approx 3 \theta_{\text{E}} + \theta_{\text{E}} \frac{(3+\pi) (|\Delta\theta_{1}| + |\Delta\theta_{2}|) \sqrt{|\Delta\theta_{1}^{2}| + |\Delta\theta_{2}^{2}|} + 2(|\theta_{\text{center},1}| + |\theta_{\text{center},2}|) \sqrt{|\Delta\theta_{1}^{2}| + |\Delta\theta_{2}^{2}|}}{|\Delta\theta_{1}|^{2} + |\Delta\theta_{2}|^{2}}  \nonumber \\
&\hphantom{=\,\,} + \theta_{\text{E}} \frac{(4.5 + \pi) |\Delta\theta_{2}|  \sqrt{|\Delta\theta_{1}^{2}| + |\Delta\theta_{2}^{2}|}}{|\Delta\theta_{1}|^{2} + |\Delta\theta_{2}|^{2}} + \theta_{\text{E}} \frac{(6+2\pi)|\Delta\theta_{1}||\Delta\theta_{2}^{2}| + 2(|\Delta\theta_{2}^{2}| + |\Delta\theta_{1}||\Delta\theta_{2}|)(|\theta_{\text{center},1}| + |\theta_{\text{center},2}|)}{(|\Delta\theta_{1}^{2}| + |\Delta\theta_{2}^{2}|)^{\frac{3}{2}}}\nonumber \\
&= \theta_{\text{E}} \bigg[3 + \frac{(3 + \pi) (|\Delta\theta_{1}| + |\Delta\theta_{2}|) + 2(|\theta_{\text{center},1}| + |\theta_{\text{center},2}|) + (4.5 + \pi)|\Delta\theta_{2}|}{\sqrt{|\Delta\theta_{1}^{2}| + |\Delta\theta_{2}^{2}|}}   \nonumber \\
&\hphantom{=\,\,} +   \frac{(6+2\pi)|\Delta\theta_{1}||\Delta\theta_{2}^{2}| + 2(|\Delta\theta_{2}^{2}| + |\Delta\theta_{1}||\Delta\theta_{2}|)(|\theta_{\text{center},1}| + |\theta_{\text{center},2}|)}{(|\Delta\theta_{1}^{2}| + |\Delta\theta_{2}^{2}|)^{\frac{3}{2}}}  \bigg].
\end{align}
We now rotate the coordinate system by -180 degrees,
\begin{align}
\nabla \Psi_{\epsilon,1} \to - \nabla \Psi_{\epsilon,1}^{\prime}, \nonumber \\
\nabla \Psi_{\epsilon,2} \to - \nabla \Psi_{\epsilon,2}^{\prime}.
\end{align}
We first compute one part of the error variable $F$, namely
\begin{align}
	|D_{1}| + |D_{2}| &= \theta_{\text{E}} \bigg[6 + \frac{(6 + 2 \pi) (|\Delta\theta_{1}| + |\Delta\theta_{2}|) + 4 (|\theta_{\text{center},1}| + |\theta_{\text{center},2}|) + (4.5 + \pi)(|\Delta\theta_{1}| + |\Delta\theta_{2}|)}{\sqrt{|\Delta\theta_{1}^{2}| + |\Delta\theta_{2}^{2}|}}   \nonumber \\
	&\hphantom{=\,\,} +   \frac{(6+2\pi)(|\Delta\theta_{1}^{2}||\Delta\theta_{2}| + |\Delta\theta_{1}||\Delta\theta_{2}^{2}|) + 2(|\Delta\theta_{1}^{2}| + |\Delta\theta_{2}^{2}| + 2 |\Delta\theta_{1}||\Delta\theta_{2}|)(|\theta_{\text{center},1}| + |\theta_{\text{center},2}|)}{(|\Delta\theta_{1}^{2}| + |\Delta\theta_{2}^{2}|)^{\frac{3}{2}}}  \bigg].\label{equation:appendix_D1_plus_D2}
\end{align}
Each term in this equation is maximized if we simultaneously maximize $\Delta\theta_{1}$ and $\Delta\theta_{2}$ for a fixed radius $\sqrt{|\Delta\theta_{1}^{2}| + |\Delta\theta_{2}^{2}|}$. We can see this by rewriting the following relations in polar coordinates, 
\begin{align}
	\frac{|\Delta\theta_{1}| + |\Delta\theta_{2}|}{\sqrt{|\Delta\theta_{1}^{2}| + |\Delta\theta_{2}^{2}|}} &= \frac{\sqrt{|\Delta\theta_{1}^{2}| + |\Delta\theta_{2}^{2}|}(|\cos(\phi^{\prime})| + |\sin(\phi^{\prime})|)}{\sqrt{|\Delta\theta_{1}^{2}| + |\Delta\theta_{2}^{2}|}} \leq \sqrt{2}, \\
	\frac{|\Delta\theta_{1}^{2}||\Delta\theta_{2}| + |\Delta\theta_{1}||\Delta\theta_{2}^{2}|}{(|\Delta\theta_{1}^{2}| + |\Delta\theta_{2}^{2}|)^{\frac{3}{2}}} &= \frac{(|\Delta\theta_{1}^{2}| + |\Delta\theta_{2}^{2}|)^{\frac{3}{2}}(\cos^{2}(\phi^{\prime})  |\sin(\phi^{\prime})|  +  |\cos(\phi^{\prime})|  \sin^{2}(\phi^{\prime}  ) )}{(|\Delta\theta_{1}^{2}| + |\Delta\theta_{2}^{2}|)^{\frac{3}{2}}} \leq \frac{1}{\sqrt{2}}, \\
	\frac{|\Delta\theta_{1}^{2}| + |\Delta\theta_{2}^{2}| + 2 |\Delta\theta_{1}||\Delta\theta_{2}|}{(|\Delta\theta_{1}^{2}| + |\Delta\theta_{2}^{2}|)^{\frac{3}{2}}} &=
	\frac{(|\Delta\theta_{1}^{2}| + |\Delta\theta_{2}^{2}|) (\cos^{2}(\phi^{\prime}) + \sin^{2}(\phi^{\prime})  +2 |\cos(\phi^{\prime})| |\sin(\phi^{\prime})|  )}{(|\Delta\theta_{1}^{2}| + |\Delta\theta_{2}^{2}|)^{\frac{3}{2}}} \leq \frac{2}{\sqrt{|\Delta\theta_{1}^{2}| + |\Delta\theta_{2}^{2}|}},
\end{align}
so we choose $\phi^{\prime} = \pi/4$. In addition, the error due to the lens center is also maximized if we maximize both center coordinates simultaneously. As a result, we have
\begin{align}
	|\Delta\theta_{1}| = |\Delta\theta_{2}|, \nonumber \\
	|\theta_{\text{center},1}| = |\theta_{\text{center},2}|.
\end{align} 
This choice also fixes the absolute value of the gradient in $\Delta \theta_{1}$ and $\Delta \theta_{2}$ direction,
\begin{align}
	|\nabla \Psi_{\epsilon,1}^{\prime}| &= |(\nabla \psi)_{1}| = \frac{\theta_{\text{E}}}{\sqrt{2}}, \nonumber \\
	|\nabla \Psi_{\epsilon,2}^{\prime}| &= |(\nabla \psi)_{2}| = \frac{\theta_{\text{E}}}{\sqrt{2}}.
\end{align}
We have used an upper limit of $\theta_{\text{E}}$ for the gradient value in the $D$ error variables which we can now replace with $\theta_{\text{E}}/\sqrt{2}$. Inserting these results into equation~\ref{equation:appendix_D1_plus_D2}, we have
\begin{align}
	|D_{1}| + |D_{2}| &= \theta_{\text{E}} \bigg[\frac{6}{\sqrt{2}} + (6 + 2\pi)\sqrt{2} + \frac{16|\theta_{\text{center},1}|}{\sqrt{2}|\Delta\theta_{1}|} + \sqrt{2}(4.5 + \pi) + \frac{6 + 2\pi}{\sqrt{2}}     \bigg] \nonumber \\
	&= \theta_{\text{E}} \bigg(16.5\sqrt{2} + 4\pi \sqrt{2} + \frac{8\sqrt{2}|\theta_{\text{center},1}|}{|\Delta\theta_{1}|}   \bigg),
\end{align}
and we obtain 
\begin{align}
F &= 2 \pi \frac{\sqrt{2}\theta_{E}}{2} + \theta_{\text{E}} \bigg(16.5\sqrt{2} + 4\pi \sqrt{2} + \frac{8\sqrt{2}|\theta_{\text{center},1}|}{|\Delta\theta_{1}|}   \bigg)  + 2 \frac{\sqrt{2}\theta_{E}}{2} \nonumber \\
 &\approx 47 \theta_{\text{E}} + \frac{8\sqrt{2}|\theta_{\text{center},1}|}{|\Delta\theta_{1}|} \theta_{\text{E}}.
\end{align}
We see that the assumption that the machine precision error is invariant under translations and rotations of the coordinate system is approximately correct far away from the lens center. The computed correction factor is less than two. However, close to the center we obtain a divergent correction term. This divergence is not a problem, as the SIE lens model itself has a divergence at the center, see e.g. equation~\ref{equation:kappa_SIS}. This unrealistic property of the SIE model is well known and other, more realistic parametric lens models do not suffer from this divergence at the center. Therefore neither single nor double precision are accurate enough at the center, but since the SIE is not a realistic lens model at its center, this is perfectly acceptable.\\
\\
First, we compute the error variable $F^{\prime}$ without the $1/|\Delta \theta_{1}|$ term for a cluster-scale halo with $\theta_{\text{E}} = 20~\text{arcsec}$,
\begin{equation}
F^{\prime}_{\text{cluster-scale}} = 940~\text{arcsec},
\end{equation}
and for a galaxy-scale halo with $\theta_{\text{E}} = 0.2~\text{arcsec}$,
\begin{equation}
F^{\prime}_{\text{galaxy-scale}} = 9.4~\text{arcsec}.
\end{equation}
For single and double precision, we have respectively $\epsilon_{\text{SP}} \approx 1.2 \times 10^{-7}$ and $\epsilon_{\text{DP}} \approx 2.2 \times 10^{-16}$, and thus the upper error bounds
\begin{align}
\epsilon_{\text{SP}} F^{\prime}_{\text{cluster-scale}} &\approx 1.1 \times 10^{-4}~\text{arcsec}, \nonumber \\
\epsilon_{\text{SP}} F^{\prime}_{\text{galaxy-scale}} &\approx 1.1 \times 10^{-6}~\text{arcsec}, \\
\epsilon_{\text{DP}} F^{\prime}_{\text{cluster-scale}} &\approx 2.1 \times 10^{-13}~\text{arcsec}, \nonumber \\
\epsilon_{\text{DP}} F^{\prime}_{\text{galaxy-scale}} &\approx 2.1 \times 10^{-15}~\text{arcsec}.
\end{align}
\\
Now we maximize the magnitude of the $1/|\Delta \theta_{1}|$ correction term by assuming a lens center 
\begin{align}
(\theta_{\text{center},1},\theta_{\text{center},2}) = (200~\text{arcsec},200~\text{arcsec}).
\end{align}
The smallest non-divergent separation from the lens center is one pixel and for a HST ACS image with a pixel size of $0.03~\text{arcsec}$ we obtain a correction
\begin{align}
	F^{\text{C}}_{\text{cluster-scale}} \approx 1.5 \times 10^{6}~\text{arcsec}, \nonumber \\
	F^{\text{C}}_{\text{galaxy-scale}} \approx 1.5 \times 10^{4}~\text{arcsec},
\end{align}
and thus
\begin{align}
\epsilon_{\text{SP}} F^{\text{C}}_{\text{cluster-scale}} &\approx 1.8 \times 10^{-1}~\text{arcsec}, \nonumber \\
\epsilon_{\text{SP}} F^{\text{C}}_{\text{galaxy-scale}} &\approx 1.8 \times 10^{-3}~\text{arcsec}, \\
\epsilon_{\text{DP}} F^{\text{C}}_{\text{cluster-scale}} &\approx 3.3 \times 10^{-10}~\text{arcsec}, \nonumber \\
\epsilon_{\text{DP}} F^{\text{C}}_{\text{galaxy-scale}} &\approx 3.3 \times 10^{-12}~\text{arcsec}.
\end{align}
The accuracy requirement computed in section~\ref{section_finite_precision_errors} shows that single precision is not accurate enough very close to the center of an isolated cluster lens, even in the absence of a magnification $M_{i}$. For an isolated galaxy lens, it is sufficient close to the center as long as $M_{i} \leq 4$. Therefore we will use double precision to compute the gradients in a pixel grid of $400 \times 400~\text{pixels}$ centered on the respective cluster lens halos and in a grid of $20 \times 20~\text{pixels}$ centered on the respective galaxy lens halos. For a HFF-like lens with 700 galaxy-scale halos and two cluster-scale halos we thus have to use double precision for $6 \times 10^{5}~\text{pixels}$ out of a total of $45 \times 10^{6}~\text{pixels}$. This corresponds to approximately 1\% of all image pixels. The correction terms for cluster-scale and galaxy-scale halos at a separation of $201~\text{pixels}$ and $11~\text{pixels}$ are respectively 
\begin{align}
	F^{\text{C}}_{\text{cluster-scale}} &\approx 7505~\text{arcsec}, \nonumber \\
	F^{\text{C}}_{\text{galaxy-scale}} &\approx 1371~\text{arcsec},
\end{align}
and thus we have 
\begin{align}
	F_{\text{cluster-scale}} &\approx 8445~\text{arcsec}, \nonumber \\
	F_{\text{galaxy-scale}} &\approx 1381~\text{arcsec},
\end{align}
and 
\begin{align}
\epsilon_{\text{SP}} F_{\text{cluster-scale}} &\approx 1.0 \times 10^{-3}~\text{arcsec}, \nonumber \\
\epsilon_{\text{SP}} F_{\text{galaxy-scale}} &\approx 1.7 \times 10^{-4}~\text{arcsec}, \\
\epsilon_{\text{DP}} F_{\text{cluster-scale}} &\approx 1.9 \times 10^{-12}~\text{arcsec}, \nonumber \\
\epsilon_{\text{DP}} F_{\text{galaxy-scale}} &\approx 3.0 \times 10^{-13}~\text{arcsec}.
\end{align}
The computed gradients for each halo are finally added up to obtain the total gradient, 
\begin{equation}
\nabla \Psi_{\epsilon,i} = \sum_{k} \nabla \Psi_{\epsilon,i,k}^{\prime},
\end{equation}
and as a result, the respective errors are combined as well. However, the respective errors can have different signs and magnitudes, so we expect to see some error cancellation. We estimate the total gradient error in the following way: We add the error contributions of two galaxy-scale lenses including upper bounds on the correction terms, but we neglect the remaining galaxy-scale halos and we add the respective upper error bounds of the cluster-scale halos. Neglecting the remaining galaxy-scale lenses is justified, because the dominating correction term decreases quickly with separation from the lens center and we expect only very few galaxies to be so close to each other that their respective correction terms are non-negligible and add up. The errors without correction term are three orders of magnitude smaller than those of the cluster-scale halos and we add many of these lenses, which are usually scattered throughout the image, so we expect significant error cancellation effects. The error contribution from the typically two cluster-scale halos will depend on their respective parameters. To obtain an upper bound, we will add up the respective upper bounds on the gradient. In total, we have
\begin{align}
\Delta (\nabla \Psi_{\epsilon,i})_{SP} &\approx 2.3 \times 10^{-3}~\text{arcsec}, \nonumber \\
\Delta (\nabla \Psi_{\epsilon,i})_{DP} &\approx 4.4 \times 10^{-12}~\text{arcsec}.
\end{align}

\end{document}